**Spontaneously Formed Orientation Polarization Thin Films for Engineering Organic-Organic Interfaces**

*Masaki Tanaka*, Rena Sugimoto, Nobuhumi Nakamura*

Department of Biotechnology and Life Science, Faculty of engineering, Tokyo University of Agriculture and Technology, 2-24-16, Naka-cho, Koganei, Tokyo, 184-8588, Japan

E-mail: m-tanaka@me.tuat.ac.jp

Abstract

Spontaneous orientation polarization (SOP) of polar molecules is formed in vacuum-deposited films by tilting their permanent dipole moment against the substrate surface direction. In this study, we developed SOP molecules with high structural asymmetry by introducing multiple fluoroalkyl groups into polar molecules because SOP is driven by asymmetric intermolecular interactions on the film surface during the vacuum deposition. The developed polar molecules exhibited high dipole orientation degrees in vacuum-deposited films and achieved a high surface potential growth rate relative to the film thickness, over −350 mV $nm^{-1}$, which is a record high for the reported compounds. Furthermore, the dipolar layers introduced at organic thin-film interfaces in hole-only devices and organic photovoltaics to study the impact of dipole interlayers on device performance. The characteristics of the device were observed to be significantly influenced by the SOP polarity, suggesting that the SOP at the organic thin-film interface plays a crucial role in charge transfer and energy level alignment. The findings of this study provide methodologies for the formation of highly anisotropic glassy films, leading to improved performance of organic devices.

## 1. Introduction

In recent years, the spontaneous orientation polarization (SOP) of organic films has attracted attention as a method for fabricating dipolar films without any polarization process after film formation.[1–3] The SOP of polar small molecules is formed through deposition via vacuum evaporation, which generates a giant surface potential (GSP) on the film surface. For example, tris(8-quinolinolato)aluminum ($Alq_3$), used in organic light-emitting diodes (OLEDs), formed SOP in a vacuum-deposited layer, and the GSP growth rate relative to the thickness (GSP slope) was approximately 50 mV $nm^{-1}$.[4,5] The SOP layers in organic semiconductor devices, such as OLEDs, affect the charge accumulation properties at the film interfaces and the quenching rate of electrically generated excitons by the accumulated charges.[6,7] Furthermore, the SOP in organic stacks drives and control the exciton dissociation and charge storage performances.[8–10] It is essential to control the SOP magnitude and polarity for these applications. In particular, the formation of a strong SOP is favored to achieve highly efficient exciton dissociation and charge storage properties. Another application of SOP films is as electret materials for vibration power generators, which also need a strong SOP film.[11,12]

SOP formation in vacuum-deposited films is driven by the surface equilibration mechanism.[13–15] Briefly, the orientation of the permanent dipole moment (PDM) is formed during surface diffusion processes during deposition, and the SOP in the film originates from the PDM orientation on the film surface. Thus, controlling the molecular diffusivity on the surface critically affects the molecular orientation in films, indicating that the film SOP depends on the substrate temperature ($T_s$) and deposition rate. Previous studies on organic glassy films revealed the significant impact of such process parameters on SOP formation; that is, a low $T_s$ and high deposition rate effectively reduce the surface diffusivity of deposited molecules and increase the PDM orientation degrees in films.[6,7,16–20] Furthermore, the codeposition of polar molecules with nonpolar host molecules, that is, the dilution of PDMs, also improves the PDM orientation in films.[17,21,22] This indicates that dipole-dipole interactions between polar molecules stabilize the antiparallel PDM orientation, reducing the SOP and PDM orientation degrees in neat films of polar molecules. Thus, to establish an SOP, it is essential to design asymmetric intermolecular interactions on the film surface during deposition to overcome dipole-dipole interactions.

The asymmetry of the intermolecular interactions on the film surface is driven by the molecular shapes and introduced functional groups. The vacuum-deposited films of $Alq_3$ and tris(7-propyl-8-hydroxyquinolinato)aluminum exhibited positive and negative GSPs, respectively.[23] This result can be explained by the difference in their molecular shapes, which stabilized the molecular orientation direction. In recent years, some techniques have been proposed for the active design of intermolecular interactions based on the introduction of functional groups with distinct properties.[17–19,24] Our group reported that controlling the polarizabilities of molecular edges can improve the molecular head-tail orientation in vacuum-deposited films.[17,18,24] The atomic polarizabilities of molecular edges govern the magnitude of dispersion forces; thus, the molecular edge with a high polarizability preferentially orients toward the substrate side on the film surface during deposition, and vice versa. The polarizabilities of the molecular edge depend on the atoms and bonding manners; for example, alkyl and fluoroalkyl groups possess relatively low polarizabilities, whereas aromatic groups and atoms, such as N and S, possess relatively high polarizabilities.[24,25] Thus, designing molecular asymmetry using functional groups with distinct polarizabilities is essential. However, the magnitude of the GSP slopes and PDM orientation degrees remain low because of the limited backbones and strategies to control asymmetry.

Surface-segregated monolayers (SSMs) are recognized as a method for the spontaneous formation of dipole layers, distinct from SOP.[26–31] During solution processes, the segregation and orientation of polar molecules at interfaces facilitate the formation of high-degree orientation polarization layers.[28] The application of SSM has facilitated the investigation of the function of dipole layers at the heterojunction interface within the active layer of organic photovoltaics (OPVs). Prior studies have indicated that the energy level shifts induced by introducing SSMs at the donor (D)/acceptor (A) molecular interface predominantly influence the open-circuit voltage ($V_{OC}$) of OPVs.[31] The $V_{OC}$ of OPVs is contingent upon the disparity between the highest occupied molecular orbital (HOMO) level of the donor and the lowest unoccupied molecular orbital (LUMO) level of the acceptor. Given that the dipole layer and the energy level structure at the interface influence this effective level difference between the donor and acceptor layers, it is feasible to modulate the $V_{OC}$. Moreover, it has been established that when SSM is intermixed at the D/A interface through the annealing of the device, the influence of the dipole layer on the device characteristics is nullified.[31] While the impact of introducing SOP at the interface of organic thin films has been extensively investigated in OLEDs,[17,18] there is a paucity of research concerning its effects at the D/A interface of OPVs. The thickness of the dipole layer formed by SOP can be precisely controlled through vacuum deposition, enabling the assessment of how variations in dipole layer thickness influence the properties of the organic thin film interface.

In this study, we demonstrate the impact of multiple introductions of orientation-inducing groups, such as $CF_3$ or long-chain fluoroalkyl groups (**Figure 1 (a)**), on SOP formation. The multiple introductions successfully improved the PDM orientation of polar molecules in vacuum-deposited films, resulting in a GSP slope of over −350 mV $nm^{-1}$. This is the highest value reported for these compounds (**Figure 1 (b) and Table S1**). The dilution effect was also demonstrated using the codeposition technique of developed polar molecules with a nonpolar molecule. The experimental results showed a significant increase in the PDM orientation degree and a nearly perfect PDM orientation in the codeposited films. In order to examine the influence of the dipole layer at organic thin-film interfaces, SOP thin films were incorporated at these interfaces within HOD and heterojunction OPVs, and their device characteristics were subsequently assessed. The findings indicated that each device displayed distinct characteristics contingent upon the polarity of the SOP. This suggests that the dipole layer plays a role in modulating charge transport and energy level structure at the organic thin-film interface, thereby enhancing the performance of organic thin-film devices. Our findings pave the way for the formation of highly functional organic films and electronic devices.

## 2. Results and discussion

We utilized 9,9-bis(3,4-dicarboxyphenyl)fluorene dianhydride as a precursor to synthesize bent-shaped SOP molecules incorporating fluorene and phthalimide (PI) units (FDI backbone), as shown in **Figure 1 (a)** (**Figures S1–8**). Our previous calculation results showed that the aromatic hydrocarbons such as fluorene possess a relatively high polarizability, tending to orient the substrate side on the film surface during deposition.[24] The FDI-based polar molecules designed in this study, that is, FDI-2TFB and FDI-2bTFB, comprise an FDI unit and $CF_3$-substituted biphenyl moieties, such as trifluoromethylbiphenyl (TFB) and bis(trifluoromethyl)biphenyl (bTFB) groups. The PDM magnitudes of FDI-2TFB and FDI-2bTFB were calculated to be 7.8 and 7.4 Debye, respectively, and the TFB/bTFB sides of the molecules were negatively polarized owing to the strong electron-withdrawing abilities of the $CF_3$ and PI units (**Figure S9**). Both FDI-based polar molecules are estimated to possess three major molecular conformations with different PDM magnitudes because of the different directions of the PIs (**Figure S9**). **Figure S10** shows the relationship between the PDM magnitude and population estimated by using CONFLEX 9 with the MMFF94s force field. The mean PDM magnitude ($\langle p \rangle$) was calculated as the weighted average of the PDM magnitude of conformers, that is, $\langle p \rangle = \sum(c_i \times p_i)$, where $p_i$ and $c_i$ are the $i$th PDM magnitude and the population of the conformers, respectively. Furthermore, the atomic polarizabilities of developed molecules were also calculated using DFT methods, and confirmed that the polarizabilities of the molecular edges with $CF_3$ units are lower than those of another edge, that is, the fluorene core (**Figure S11-17 and Table S2-8**). The difference in intramolecular polarizabilities will be the driving force to form asymmetric molecular orientation on the surface during the vacuum deposition.[24]

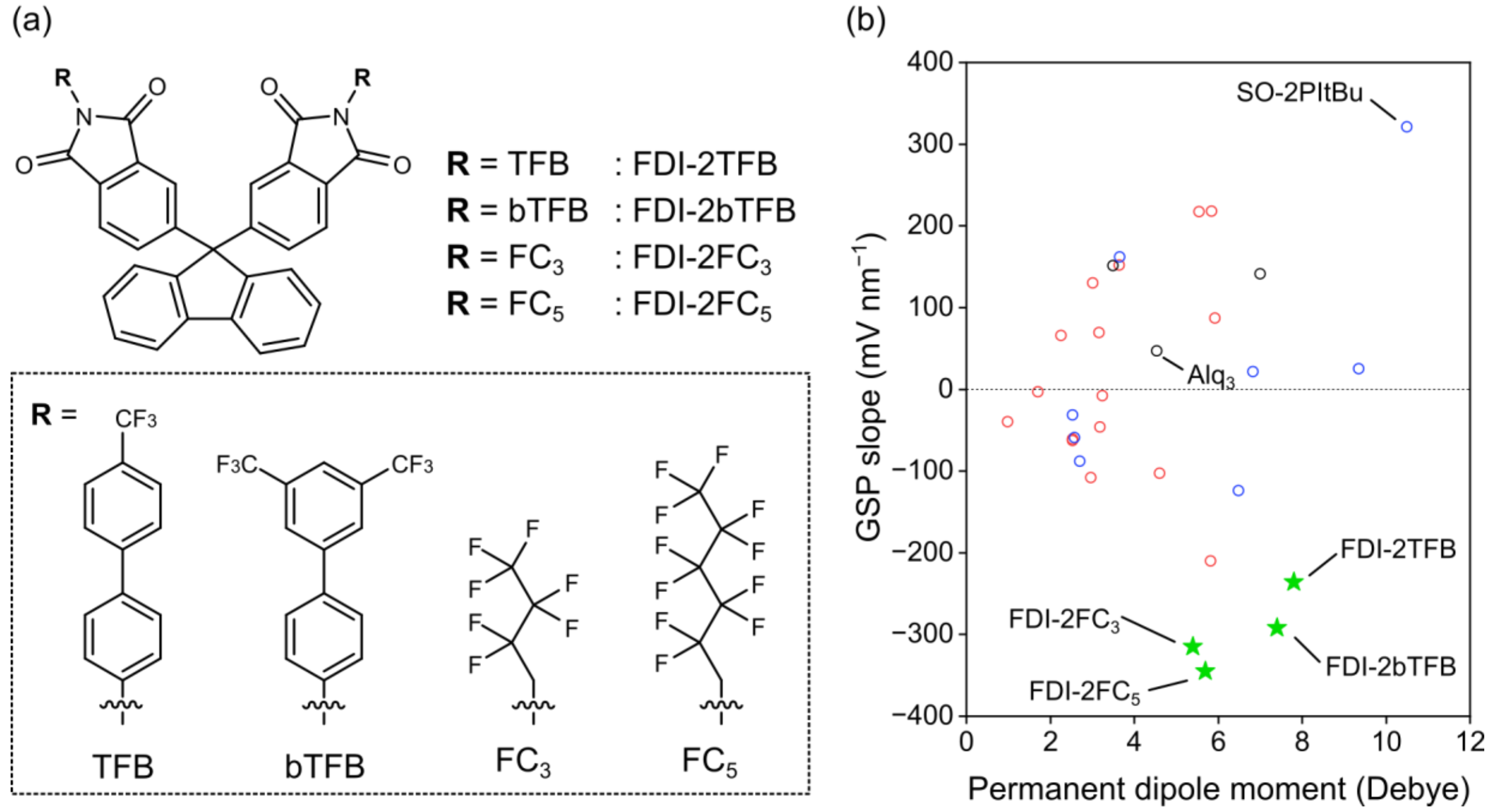

**Figure 1** Molecular design of the FDI-based SOP molecules. (a) Chemical structures of polar molecules developed in this study. (b) Relationship between permanent dipole moment and the reported giant surface potential (GSP) slope values of vacuum-deposited films.[3,4,17-19,24] Black symbols represent the values of typical reported polar molecules.[3,4,19] Red symbols represent the slope values of previously reported F-containing molecules, and blue symbols denote those of molecules without F-based units developed by the authors' group.[17,18,24] Green star symbols represent the slopes of the molecules developed in this study.

Vacuum-deposited films of FDI-2TFB and FDI-2bTFB were formed on an indium tin oxide (ITO)-coated glass substrate using a physical vapor deposition technique. The surface potential of the deposited organic films was measured using a Kelvin probe by transferring the substrate to the measurement chamber without breaking the vacuum after film preparation. We investigated the thickness dependence of the GSPs of deposited organic films to estimate the GSP growth rate with respect to the film thickness (GSP slope), which is an indicator of the SOP magnitude. **Figure 2 (a)** shows the thickness dependence of the GSPs of the FDI-2TFB and FDI-2bTFB films deposited at a rate of approximately 0.1 nm $s^{-1}$. Both films exhibited negative GSPs, and their magnitudes increased with thickness, indicating that their molecular PDMs exhibited spontaneous orientation. The GSP slopes were determined to be −236 and −292 mV $nm^{-1}$ for FDI-2TFB and FDI-2bTFB, respectively. The negative GSP values clearly indicate the spontaneous orientation of polar molecules, in which the $CF_3$ groups in the molecules preferentially orient toward the vacuum (V-type orientation shown in **Figure S18**) on the film surface during vacuum deposition because of the low polarizability of the $CF_3$ unis.[17,18] The degree of PDM orientation $\langle\cos\theta\rangle$ of polar molecules in the films was calculated as $\langle\cos\theta\rangle = m\varepsilon_r\varepsilon_0/\langle p\rangle n$, where $\theta$, $m$, $\varepsilon_r$, $\varepsilon_0$, and $n$ are the tilt angle of the molecular PDM with respect to the substrate normal, GSP slope, relative permittivity, dielectric constant of vacuum, and number density in the films, respectively. The values of $\varepsilon_r$ were determined using capacitance ($C$) measurement of sandwiched device, that is, ITO (100 nm)/organic film ($d$ nm)/Al (100 nm) without charge injection conditions. The frequency dependence of device capacitance ($C$-$f$) was measured the frequency range from $10^5$ to $10^2$ Hz (**Figure S19**). The obtained $C$ were almost constant in tested frequency range, so the $C$ values were averaged over the frequency range. The $\varepsilon_r$ values (**Table S9**) were calculated as $\varepsilon_r = Cd/\varepsilon_0 S$, where $d$ and $S$ are the organic film thickness and device area, respectively. The values of $n$ were calculated as $n = \rho N_A/M$, where $\rho$, $N_A$, and $M$ are the film density, Avogadro's number, and molecular weight, respectively. The $\rho$ values of the vacuum-deposited films were estimated using a well-calibrated quartz crystal oscillator and the previously reported $\rho$ value of another polar molecule (SO-2PItBu).[24] The values of $\rho$ and $n$ are listed in **Table S10**. The $\langle\cos\theta\rangle$ values of FDI-2TFB and FDI-2bTFB were −0.27 and −0.32, respectively. The higher $|\langle\cos\theta\rangle|$ indicates that the ordered PDM orientation can be achieved by the multiple introduced $CF_3$-units, which effectively lower the polarizability of the functional groups at the edge side of the molecules.

We performed complimentary studies to clarify the effect of end-group on the orientation degree of polar molecules (**Figure S20 and Table S11**). Additionally developed FDI-2*p*TFP and FDI-2*m*TFP are the positional isomers each other, and their GSP slopes and orientation degrees were comparable. This result indicates that reducing polarization at the molecular end is more important for molecular orientation than the substitution position of the $CF_3$ group in the polar molecules. Another molecule, FDI-2mTFPCN, with additional polar cyano groups at the molecular ends has a large PDM magnitude of 11.7 Debye. However, the GSP slope of FDI-2*m*TFPCN was lower than that of FDI-2*m*TFP due to the low orientation degree of −0.028. This is attributed to the reduction of the intramolecular difference in the polarizabilities of the molecular ends.[24] We confirmed that the introduced cyano groups raised the polarizabilities of the molecular ends, leading to the reducing asymmetry of intermolecular van der Waals interactions for the SOP formation. These results also indicate that the controlling polarizability is one of the important issues to achieve the formation of highly ordered molecular orientation.

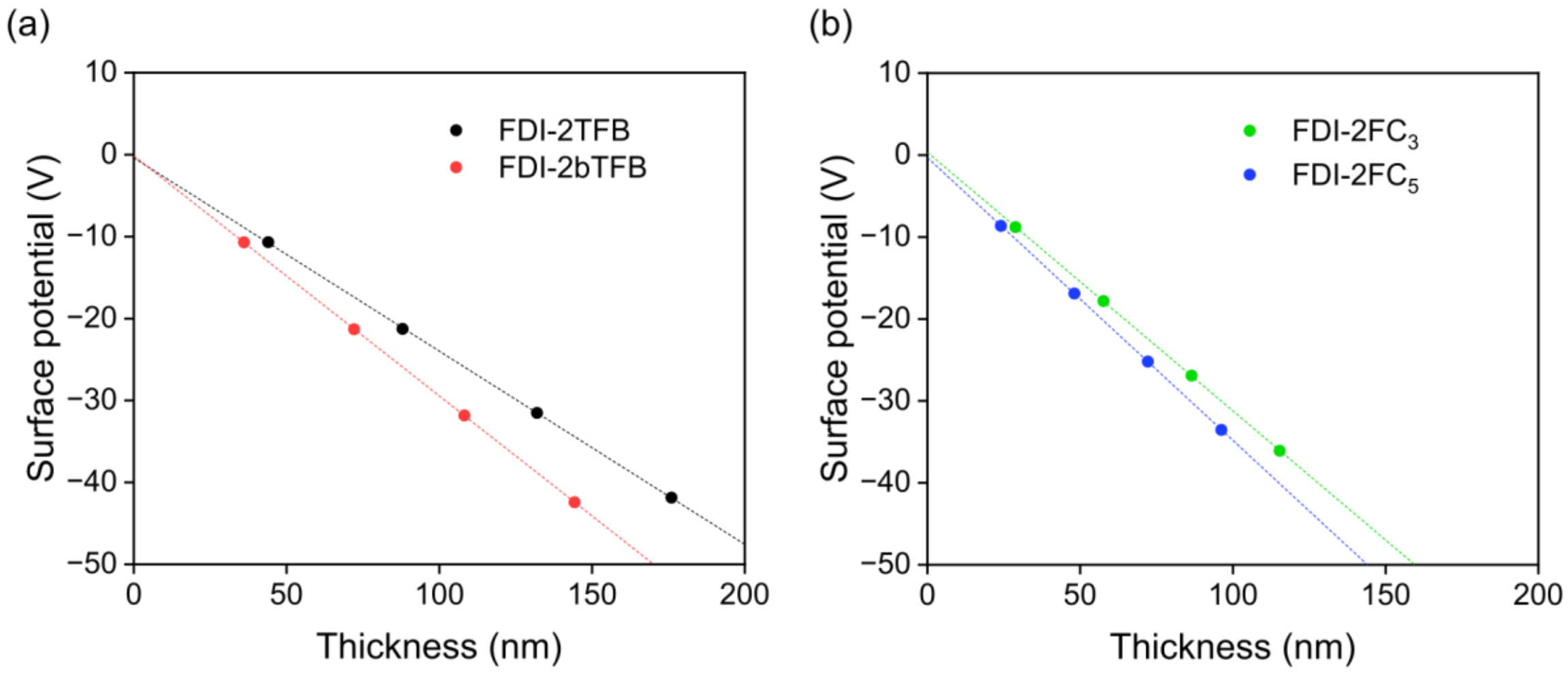

**Figure 2** Thickness dependence of the surface potential of the developed polar molecules. (a) FDI-2TFB and FDI-2bTFB. (b) FDI-2FC$_3$ and FDI-2FC$_5$.

An alternative approach to reduce the polarizability of the molecular ends is to introduce long-chain fluoroalkyl groups instead of TFB groups. We designed FDI-based polar molecules, such as FDI-2FC$_3$ and FDI-2FC$_5$, with heptafluorobutyl (FC$_3$) and undecafluorohexyl (FC$_5$) moieties at the edges, which facilitate asymmetric polarizability to weaken molecular interactions and reinforce anisotropic molecular orientations. The <*p*> values of FDI-2FC$_3$ and FDI-2FC$_5$ were estimated to be 5.4 and 5.7 Debye, respectively. **Figure 2 (b)** shows the thickness dependence of the GSPs of the FDI-2FC$_3$ and FDI-2FC$_5$ films. The GSP slopes were determined to be −315 and −345 mV nm$^{-1}$ for FDI-2FC$_3$ and FDI-2FC$_5$, respectively. Although these thin films showed negative GSPs, similar to FDI-2TFB and FDI-2bTFB, the GSP slopes were higher than those of the TFB-based polar molecules. The <cos*θ*> values of FDI-2FC$_3$ and FDI-2FC$_5$ were calculated to be −0.29 and −0.37, respectively. These results clearly indicate that the introduction of long-chain fluoroalkyl groups also induces efficient SOP owing to the highly asymmetric molecular structures. The stability of SOP was also tested. A sample structure for stability test was FDI-2FC$_5$ (85 nm)/ 4,4',4''-tri-9-carbazolyltriphenylamine (20 nm)/ITO (100 nm). The interlayer of 4,4',4''-tri-9-carbazolyltriphenylamine (TCTA) was used to suppress charge injections to the SOP layer from the ITO electrode.[32] The SOP film sample was stored in the atmosphere under dark conditions. Temperature and relative humidity were maintained at 22 ± 2 °C and 35±10%, respectively. **Figure S21** shows the stored time dependence of the surface potentials of a film sample, and the surface potential decreased by approximately 15% from its initial value over a 30-day period. This indicates that although FDI-2FC$_5$ film is an SOP film with relatively high orientation degree, its orientation state is fundamentally stable.

**Table 1 Summary of molecular and film properties of the studied compounds.**

| | $\langle p \rangle$[a] (Debye) | GSP slope (mV nm$^{-1}$) | Mean orientation degree | $T_g$[b] (°C) |
|---|---|---|---|---|
| FDI-2TFB | 7.8 | −236 | −0.27 | 142 |
| FDI-2bTFB | 7.4 | −292 | −0.32 | 167 |
| FDI-2FC$_3$ | 5.4 | −315 | −0.29 | 97 |
| FDI-2FC$_5$ | 5.7 | −345 | −0.37 | 86 |

[a] Mean permanent dipole moment magnitude

[b] Glass transition temperature

It has been reported that a longer diffusion time (high molecular diffusivity) to relax molecular orientations results in the formation of more stabilized orientation states, that is, random or antiparallel PDM orientation states, which induce small or zero SOP magnitude in films.[14,16,19] The diffusion time is directly controlled by tuning the deposition rate, which is the origin of the deposition rate dependence of GSP. **Figure 3** shows the thickness dependence of the GSPs of the FDI-2FC$_5$ films deposited at various deposition rates and the deposition rate dependence of the GSP slopes. The GSP slopes of FDI-2FC$_5$ exhibited a monotonic positive dependence on the deposition rates and changed from −292 to −356 mV nm$^{-1}$ in the range of deposition rates from 0.017 to 0.18 nm s$^{-1}$. These results indicate that the reduction in the surface diffusion time improved the degree of PDM orientation from −0.31 to −0.38 in this range of deposition rates, assuming a constant film density and permittivity in the range of the tested deposition rates. However, FDI-2bTFB did not show a monotonic dependence of the GSP slopes on the deposition rates, but convex-shaped trend (**Figure S22**). That is, the GSP slope of FDI-2bTFB showed small values at low and high deposition rates and the highest values at the deposition rate of 0.05-0.1 nm $s^{-1}$. This deposition rate dependence indicates that the degree of the PDM orientation FDI-2bTFB is lowered at an excessively high deposition rate. We estimate that these differences in the deposition rate dependences between the two polar molecules are attributed to the intrinsic factor correlating the surface diffusion time or diffusivity. One of the indicators for surface diffusivity in the research area in organic glassy materials is the ratio of $T_s$ and glass transition temperature ($T_g$), $T_s/T_g$.[33–36] The $T_g$ values of the materials were measured using differential scanning calorimetry (**Figure S23**), and the values are listed in **Table 1**. The $T_g$ of FDI-2bTFB (167 °C) was significantly higher than that of FDI-2FC$_5$ (86 °C), indicating that FDI-2bTFB has a shorter surface diffusion time compared to FDI-2FC$_5$. A short diffusion time improves PDM orientation degrees due to the suppression of the formation of random/anti-parallel dipole orientations in deposited films as shown in the rate dependence of FDI-2FC$_5$'s GSP slope. However, when a diffusion time is shorter than that of molecular reorientation to form the orientation polarization at the surface after the adsorption of molecules, the PDM orientation would be random. Thus, the orientation degree of FDI-2bTFB with high $T_g$ (low $T_s/T_g$) showed a decrease at the high deposition rate leading to the limitation of molecular surface diffusion, and the highest values at the balanced deposition rates based on the surface diffusion time and reorientation time to form the orientation. The reduction of orientation under conditions of suppressed molecular diffusion has also been reported in papers by Cakaj et al.[19] In general, a decrease in $T_s/T_g$ increases the degree of orientation, but they observe that very low $T_s/T_g$ decrease orientation of bis-4-(N-carbazol-yl-phenyl)phenyl phosphine oxide (BCPO) because of the too slow diffusion to find suitable sites for favorable orientation. In this study, to investigate this phenomenon, the dependence of the GSP slope of the FDI-2bTFB films on the deposition rate on a heated substrate was examined. The GSP slopes of the FDI-2bTFB films deposited on an ITO substrate heated at $T_s$ = 74 °C ($T_s/T_g$ = 0.79) exhibited monotonic increase with the deposition rates (**Figure S22**); that is, typical deposition rate dependence such as an FDI-2FC$_5$ case. These results clearly indicate that the deposition rate dependence of the PDM orientation originates from a change in the surface diffusion time, and there is the optimal deposition rate for PDM orientation formation depending on the $T_s/T_g$ value. We note that we estimate that the reason for the monotonic and typical dependence of GSP slopes of polar molecules, such as FDI-2FC$_5$ and BCPO, etc.[16], would be lower $T_g$ and only the limited deposition rate was tested. Therefore, it is necessary to investigate the wide range of deposition rates for complete understanding of the dynamics of SOP formation.[37]

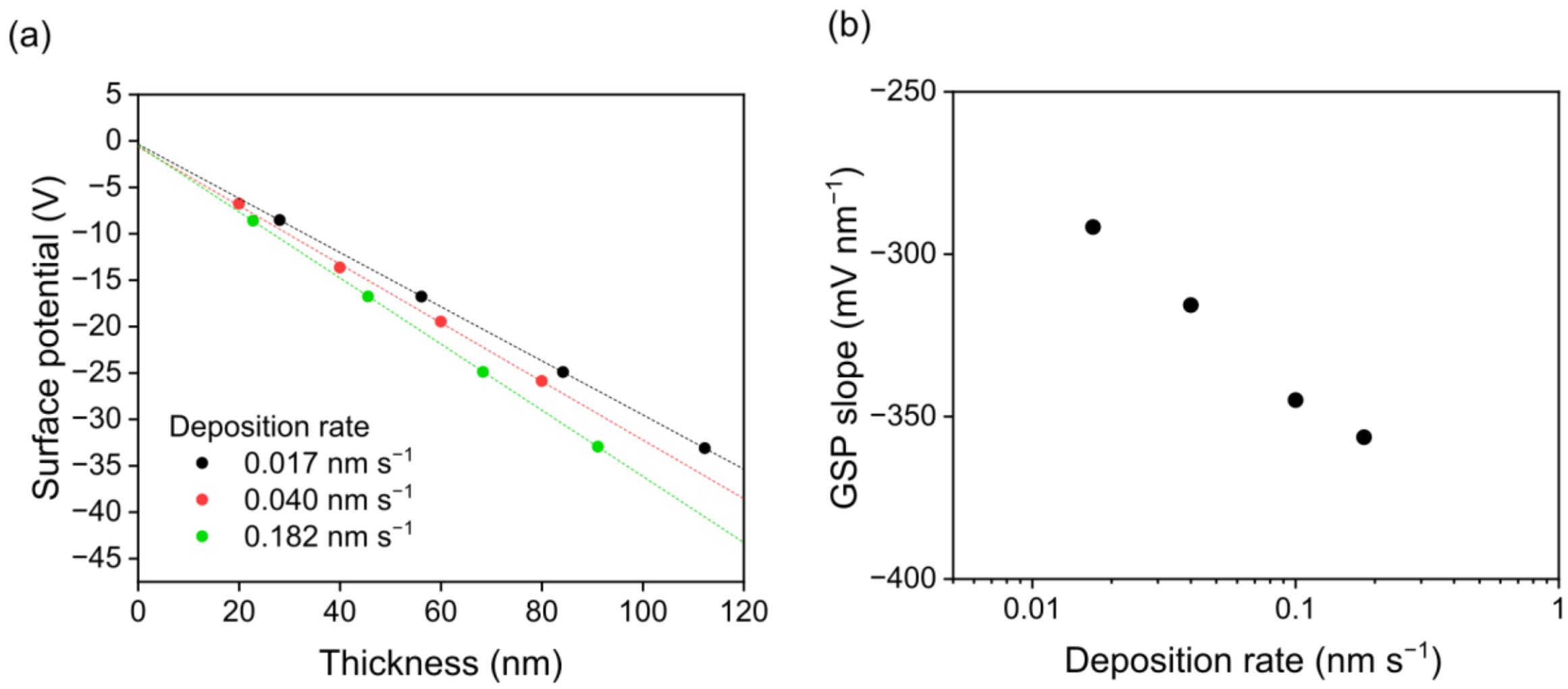

**Figure 3** Impact of deposition rate on spontaneous orientation polarization. (a) Thickness dependence of the surface potentials of the FDI-2FC$_5$ films deposited at various deposition rates. (b) Deposition rate dependence of the GSP slope of the FDI-2FC$_5$ films.

The codeposition of SOP molecules with a nonpolar host is also a technique for controlling the PDM orientation in films. This is attributed to the suppression of intermolecular dipole-dipole interactions between the SOP molecules by the nonpolar host molecules, which improves the PDM orientation owing to the stabilized parallel PDM orientations[17,21,22] We examined the SOP properties of codeposited films based on the developed FDI-FC$_5$ and 2-(9,9'-spirobi[fluoren]-3-yl)-4,6-diphenyl-1,3,5-triazine (SF3-TRZ).[38] We note that SF3-TRZ can be considered a nonpolar molecule, and its SOP is negligible (**Figure S24**, GSP slope = 3.8 mV nm$^{-1}$). **Figure 4 (a) and (b)** show the thickness dependence of the surface potential and the concentration dependence of the GSP slope of the FDI-2FC$_5$ codeposited films, respectively. The GSP slopes of the codeposited films monotonically decreased with decreasing FDI-2FC$_5$ concentration. In contrast, the calculated orientation degrees of FDI-2FC$_5$ monotonically increased and reached approximately −0.99, that is, nearly perfect PDM orientation in the codeposited film with 8 mol%-doped FDI-FC$_5$. This indicates that if we can perfectly reduce the intermolecular dipole-dipole interactions in a neat FDI-2FC$_5$ film, nearly 100% PDM orientation can be achieved owing to the well-designed molecular asymmetry. To achieve a higher PDM orientation, the molecular design for the suppression of dipole-dipole interactions is critically essential in the future.

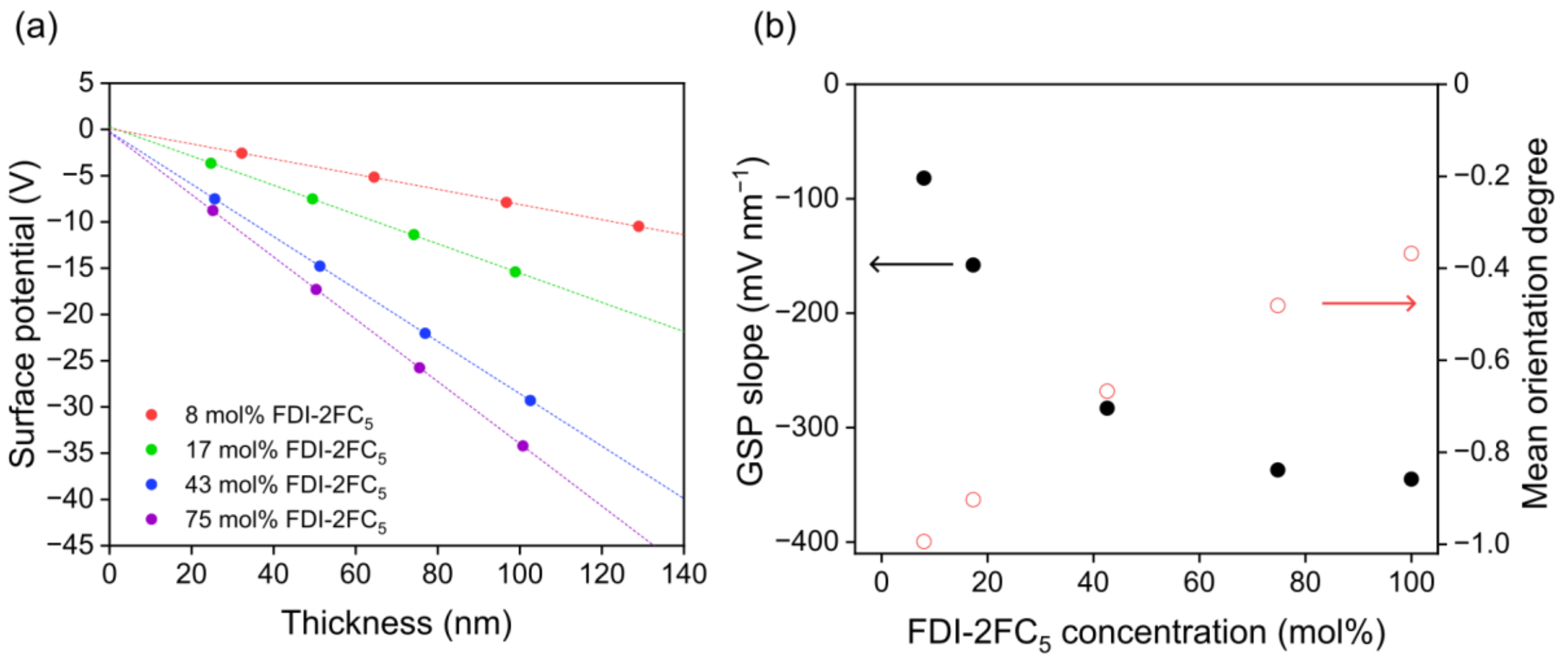

**Figure 4** Codeposition of SOP molecules with nonpolar host molecules. (a) Thickness dependence of the surface potentials of FDI-2FC$_5$ codeposited films with SF-TRZ at various FDI-FC$_5$ concentrations. (b) FDI-2FC$_5$ concentration dependence of the GSP slope and mean orientation degree in the codeposited films of FDI-2FC$_5$ and SF3-TRZ.

SOP molecules provide a simplified technique for polarization formation at electrode or organic film interfaces without any post-polarization process. A previous study revealed that the introduction of an SOP layer at the organic/organic interface in OLEDs modulates the charge injection between the interfaces, which is attributed to the vacuum level shift induced by the SOP layer.[18] The stronger polarization of the SOP films (higher GSP slope) results in larger shifts in the vacuum level, lowering/raising the energy levels at the interface. Here, we examined the impact of the developed SOP molecules on the rectification properties of charge transport in hole-only devices (HODs) with a stacked structure using 4,4',4''-tris[phenyl(m-tolyl)amino]triphenylamine (m-MTDATA), molybdenum oxide ($MoO_x$), and SOP molecules, such as ITO (100 nm)/$MoO_x$ (10 nm)/m-MTDATA (50 nm)/SOP layer (2 nm)/m-MTDATA (50 nm)/$MoO_x$ (10 nm)/Al (100 nm) (**Figure S25**). **Figure 5 (a)** shows the current density-voltage characteristics of the fabricated HODs. The reference HOD without the SOP interlayer exhibited comparable current densities under positive and negative biases owing to the use of $MoO_x$ layers as a hole injection layer. Positive and negative biases indicate hole transport from the ITO- and Al sides, respectively. The HOD with an FDI-2FC$_5$ interlayer at the m-MTDATA interface exhibited rectification properties; that is, the current densities under positive bias were higher than those under negative bias. Note that we confirmed that the similar rectification properties of the HOD based on 4,4',4''-tri-9-carbazolyltriphenylamine (TCTA) used as other hole transporting layers (**Figure S25**). Furthermore, we fabricated an HOD using a previously reported positive SOP molecule, SO-2PItBu, as the SOP interlayer.[24] The rectification polarity of the SO-2PItBu-based HOD was inverted to that of the FDI-2FC$_5$-based HOD, indicating that the rectifications were tuned by the SOP polarity (**Figures S26 (a) and (b)**). The findings indicate a trend analogous to experiments where SSM was incorporated at organic thin-film interfaces, strongly implying that the presence of a dipole layer at the organic thin-film interface influences charge transport.[26]

To understand the rectification properties of the stacked structure, that is, m-MTDATA (50 nm)/SOP layer (2 nm)/m-MTDATA (50 nm), the surface potentials of the stacked films were measured using KP. The deposited film of m-MTDATA also exhibited SOP with a GSP slope of +28 mV nm$^{-1}$. The surface potential profiles of the films are shown in **Figure 5 (b)**. The FDI-2FC$_5$ film was deposited in 1-nm-thick steps on the m-MTDATA film, and the surface potential steeply decreased owing to the negative SOP of FDI-2FC$_5$. Furthermore, the m-MTDATA films deposited on FDI-2FC$_5$ showed surface potential profiles comparable to those of the m-MTDATA film on an ITO substrate. Thus, the negative SOP of the introduced FDI-2FC$_5$ was maintained at the interface in the stacked film. In the case of the SO-2PItBu interlayer (**Figure S26 (c)**), a positive SOP was formed at the m-MTDATA interface, the polarity of which was inverted with the FDI-2FC$_5$ interlayer. The HOMO and LUMO levels of the SOP molecules were determined using photoemission yield spectroscopy, and the optical gaps were estimated using the film absorption spectra (**Figure S27 and Table S12**). **Figure S28** shows the HOMO-LUMO energy diagrams of the stacked systems with SOP interlayers, including vacuum level shifts using the measured surface potentials profiles. The rectification properties of can be attributed to the energy level shift at the interface in the HODs. The negative electrical polarization between the m-MTDATA layers improved the hole transfer from the lower and upper m-MTDATA layers under positive bias application, and vice versa. In contrast, positive polarization at the interface enhances interfacial hole transfer under a negative bias.

Further investigation into the actual energy diagrams of the devices is essential to elucidate the precise mechanism underlying SOP-induced device rectification. In organic electronic devices characterized by a sandwiched configuration of organic active layers and electrodes, the interaction between the organic layers and the electrode facilitates charge injection, thereby achieving thermal equilibrium. However, numerous studies on SOP-related device analysis have demonstrated that SOP influences exciton dissociation and charge injection/accumulation in OLEDs and OPVs, indicating that SOP screening is not entirely effective, and thus, SOP persists in electronic devices.[1,2,6,32,39,40] In addressing this issue, we investigated the dependence of HOD performance on interlayer thickness. The HOD utilizing an FDI-2FC5 interlayer with a thickness of 1 nm also demonstrated current rectification. However, the rectification ratio, defined as the ratio of current densities under positive and negative biases, increased with interlayer thickness (**Figure S29**). This finding suggests that the magnitude of the energy level shift induced by the SOP interfacial layer in the device is contingent upon its thickness and functions as an interfacial charge transport barrier under negative biases.

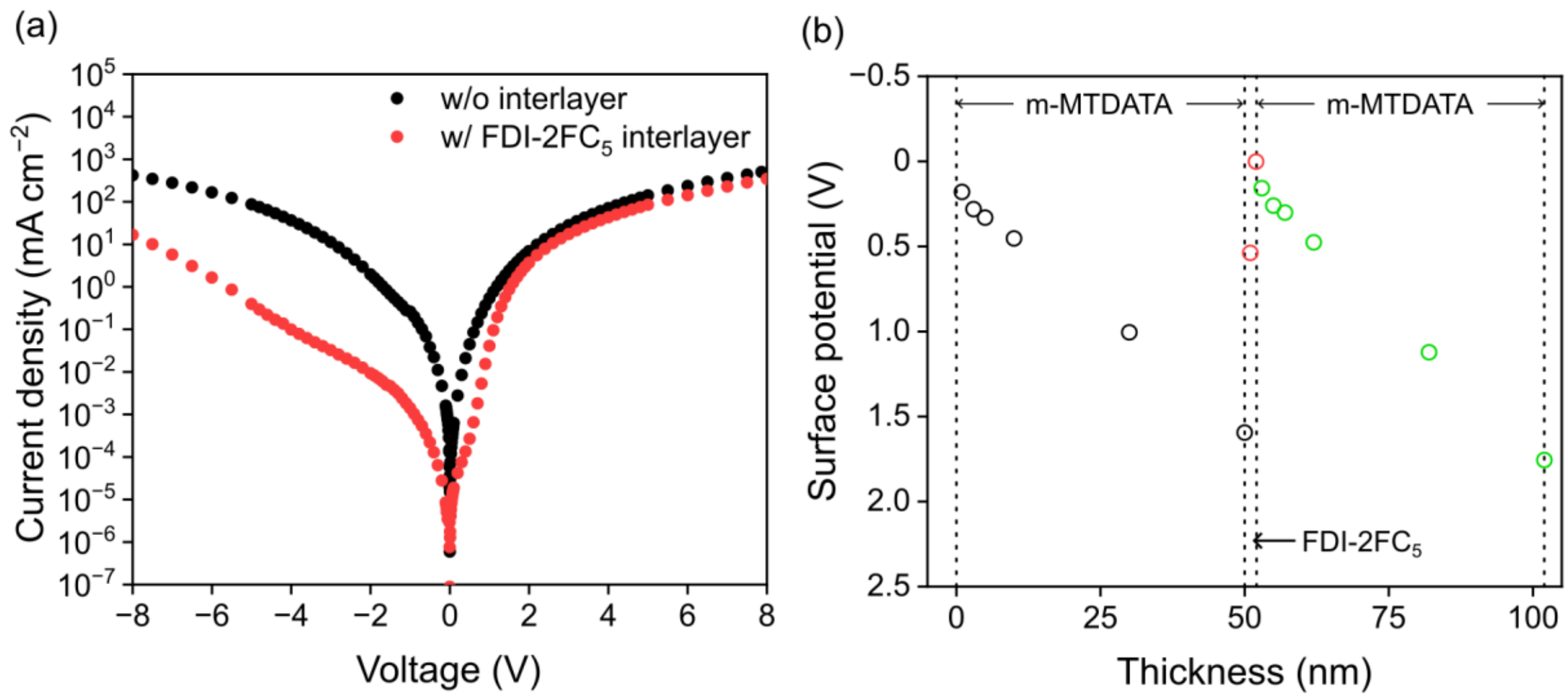

**Figure 5** Rectification properties of hole-only devices (HODs) with SOP interlayers. (a) Current density-voltage characteristics of hole-only devices with and without an FDI-2FC5 interlayer. (b) Thickness dependence of the surface potential of the m-MTDATA/FDI-2FC5/m-MTDATA stacks on an indium tin oxide (ITO) substrate.

Subsequently, we employed the synthesized SOP molecules, characterized by high GSP slopes, in planar heterojunction-based OPVs (PHJ-OPVs). At the heterojunction between the donor and acceptor layers, photogenerated excitons are dissociated into free charges via a charge-transfer state. In this context, we introduced SOP layers at the interfaces between the donor and acceptor layers to examine the influence of the SOP layer on the photovoltaic properties of PHJ-OPVs. The device structure used in this study was ITO (100 nm)/$MoO_x$ (10 nm)/m-MTDATA (50 nm)/SOP interlayer (x nm)/$C_{60}$ (50 nm)/LiF (1 nm)/Al (150 nm), where m-MTDATA and $C_{60}$ were used as donor and acceptor molecules, respectively (**Figure 6 (a) and (b)**). At first, we measured the surface potential profiles of organic film stacks deposited on a $MoO_x$/ITO substrate (**Figure S30**). The $C_{60}$ film deposited on the m-MTDATA layer exhibited a pronounced alteration in surface potential at the D/A interface (**Figure S30 (a)**). A similar pattern was observed in organic stacks incorporating a 5 nm thick SO-2PItBu interlayer (**Figure S30 (b)**). Conversely, the surface potential of the $C_{60}$ layer on the FDI-2FC$_5$ interlayer demonstrated a gradual change with respect to film thickness (**Figure S30 (c)**). Additionally, the vacuum level shift and HOMO-LUMO diagrams are presented in **Figure S30**. These diagrams were derived solely from the results of the surface potential measurements; consequently, the alignment of the Fermi level in the devices is not represented in these diagrams. Within the surface potential measurement time range, the surface potential of SO-2PItBu layer (5 nm) was maintained, despite the Fermi level crossing the LUMO level of this layer. The SOP in this layer will be partially canceled within the device, but a certain degree of polarization will be maintained. To determine the extent of the remaining polarization, a numerical calculation will be necessary in the future study.[32] On the other hand, in the case of FDI-2FC$_5$, since HOMO and LUMO levels contain the Fermi level, the screening of polarization is not so significant, and it is expected that polarization can be maintained even within the device.

The OPVs were irradiated with 365-nm UV light through the ITO electrode. The UV irradiation was predominantly absorbed by the m-MTDATA and $C_{60}$ layers because of the wide bandgaps of the SOP layers, such as FDI-2FC$_5$ and SO-2PItBu, used in this study. **Figure 6 (c) and (d)** shows the current density-voltage profiles of OPVs based on SO-2PItBu and FDI-2FC$_5$ interlayers, respectively. Although the device with simple PHJ of m-MTDATA and $C_{60}$ exhibited small $J_{SC}$ and $V_{OC}$ values, the photovoltaic performances were improved in both cases of the introduction of interlayers with different SOP polarities. However, the impact of interlayer introduction on device performance was different each other (**Figure 6 (e) and (f)**). Although introducing an SO-2PItBu layer improved both $J_{SC}$ and $V_{OC}$, the $V_{OC}$ was mainly improved by the introduction of an FDI-2FC$_5$ layer. The LUMO levels of the interlayers can establish a cascade-type energy level at the interface due to their intermediate LUMO levels between the donor and acceptor layers, thereby facilitating efficient electron transfer from m-MTDATA to $C_{60}$. We note that the HOMO levels of the interlayers are significantly deep compared to the $C_{60}$ layer. Thus, the photogenerated excitons in the m-MTDATA layer mainly contribute to the photovoltaic performance of the OPVs with interlayers. In the point of view of the cascade-type LUMO levels at the interface, both interlayers can be expected to improve charge transfer efficiency, leading to a high $J_{SC}$, however, FDI-2FC$_5$ interlayers exhibited a minimal impact on the $J_{SC}$. The variation in the impact of SOP molecules on $J_{SC}$ suggests that the polarity of SOP influences the charge transfer process at the interfaces. The positive polarity of the electric field in the SO-2PItBu interlayer can effectively facilitate the extraction of electrons from m-MTDATA excitons, thereby enhancing charge transfer efficiency and reducing charge recombination at the interface. Conversely, the negative electric field induced by FDI-2FC$_5$ is detrimental to electron transfer from donor layers, increasing charge recombination loss at the interface and consequently diminishing the photovoltaic properties.

The increase in the $V_{OC}$ of devices incorporating FDI-2FC$_5$ interlayers was noted, despite a modest rise in the $J_{SC}$. We attribute this phenomenon to the energy level shift induced by the interfacial SOP.[31] As shown in previous sections, interfacial SOP layers can form energy level shift to alter energy diagrams of heterojunction interfaces. The SOP of FDI-2FC$_5$ induces the positive shift of vacuum level at the interface and changes the alignment of energy levels of donor and acceptor layers to increase energy difference between the HOMO level of a donor layer and the LUMO level of an acceptor layer. The substantial energy disparity at the interface leads to a significant $V_{OC}$, even in the presence of a relatively small $J_{SC}$. To illustrate the impact of SOP interlayers on the polarity effect of $V_{OC}$, we compare the *J*-*V* profiles of devices incorporating interlayers under comparable $J_{SC}$ conditions (**Figure S31**). The results indicate that the use of FDI-2FC5 interlayers enhances $V_{OC}$ in comparison to SO-2PItBu interlayers. Given the critical importance of maintaining the electric field within the device for the impact of the dipolar layer on the $V_{OC}$, it is imperative to elucidate these aspects in future research.

In this study, the dipolar layer demonstrated positive effects even in the interlayer thickness range of several nanometers, which is thicker than those used in previous studies.[41,42] In fact, as an additional experiment, we fabricated and evaluated devices using 4,6-bis(3,5-di(pyridin-4-yl)phenyl)-2-methylpyrimidine (B4PyMPM) as the acceptor molecule. When the thickness of the interfacial layer was 2 nm, there was a significant decrease in $J_{SC}$ (**Figure S32**). This suggests that the interfacial layer, by increasing the separation between D and A molecules and limiting charge transfer, does not exhibit a beneficial function when the interlayer is of substantial thickness. In $C_{60}$-based devices, it is posited that the enhancement in photovoltaic performance, even with relatively substantial film thicknesses, can be attributed to the intermixing effects of deposited $C_{60}$ molecules at the interface with SOP underlayers.[41,43] The intermixing of $C_{60}$ molecules with SOP layers reduces the effective D/A separation distance, enabling efficient charge transfer through the interlayers with the thickness of several nanometers. Charge transfers between D and A layers also depend on the HOMO and LUMO levels of interlayers. The SOP molecules employed in this study facilitate electron transfer from donor D to acceptor A molecules. However, the transfer of holes from A to D is impeded due to the deep HOMO levels of the SOP molecules. To further enhance the charge transfer process, it is imperative to adjust the energy levels of SOP molecules and/or optimize the combination of D, A, and SOP molecules. Additionally, the primary exciton generation layer in this study is m-MTDATA, which demonstrates a limited light absorption band. The integration of organic dye molecules or hybrid perovskites, which possess a broader absorption band, in conjunction with an SOP interlayer, may improve device performance. The findings of this study suggest that SOP interfacial layers with varying polarities each enhance photovoltaic properties through distinct mechanisms. The incorporation of SOP molecules is essential as a novel design guideline for multilayered photovoltaic devices.

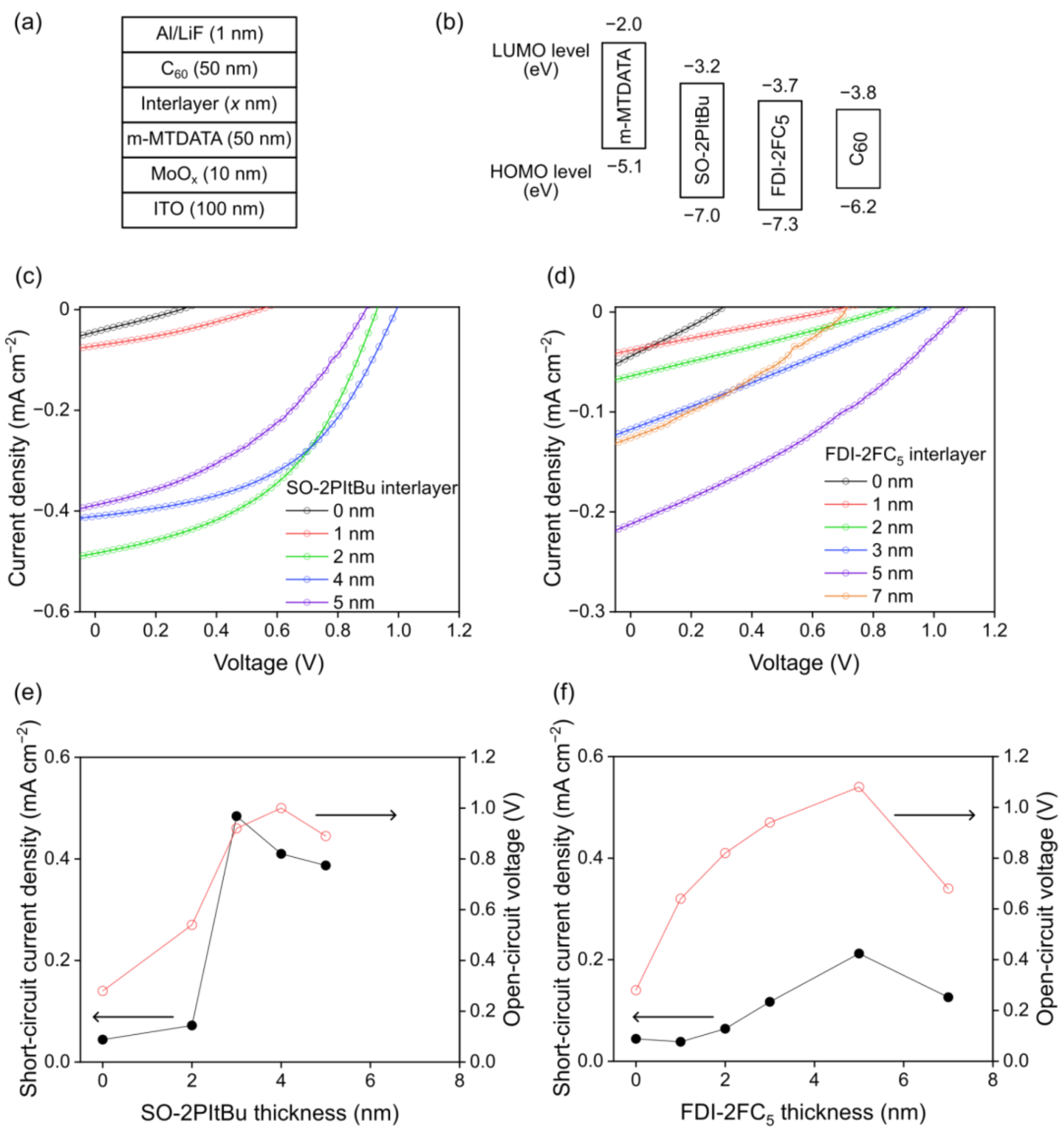

**Figure 6** Device properties of organic photovoltaic devices with SOP interlayers. (a) Device structure. (b) Energy diagrams. (c) Current density-voltage characteristics of photovoltaic devices with SO-2PItBu interlayers. (d) Current density-voltage characteristics of photovoltaic devices with FDI-2FC$_5$ interlayers. (e) Thickness dependence of the short-circuit current

density and the open-circuit voltage of devices with SO-2PItBu devices. (f) Thickness dependence of the short-circuit current density and the open-circuit voltage of devices with FDI-2FC$_5$ devices. In this study, photovoltaic devices were irradiated by 365-nm UV light with the intensity of 13.7 mW cm$^{-2}$.

In conclusion, we developed polar molecules based on the FDI core and F-containing functional groups to demonstrate the effect of multiple introductions of orientation-inducing functional groups on the SOP magnitude. The developed polar molecules exhibited a high PDM orientation degree of over 0.35 and a high GSP slope of over 350 mV nm$^{-1}$, which is a record high for the reported SOP compounds. Highly polarized SOP films generated the rectification properties for charge transport in HODs, indicating that SOP can improve the tunability of charge injection and confinement in organic electronic devices. Moreover, the photovoltaic properties of PHJ-OPVs were enhanced through the application of SOP interlayers at the donor/acceptor heterojunction. Our findings indicate that both the positive and negative polarities of SOP contribute to the improvement of photovoltaic performance through distinct mechanisms. The findings of this study provide methodologies for the formation of highly anisotropic glassy films and their applicability to devices.

**Acknowledgements**

This work was partially supported by JST FOREST Program (JPMJFR223S), JSPS KAKENHI (23K13716 and 25K01842), Advanced Technology Institute Research Grants, Asahi Glass Foundation, The Kao Foundation for Arts and Sciences. The authors thank Prof. Takahiro Ichikawa of Tokyo University of Agriculture and Technology for experimental assistance. The NMR measurements were performed at Tokyo University of Agriculture and Technology for Research Center for Science and Technology. We thank Prof. Keiichi Noguchi and the members of Research Center for Science and Technology of Tokyo University of Agriculture and Technology for their technical assistance.

# Supporting Information

Spontaneously Formed Orientation Polarization Thin Films for Engineering Organic-Organic Interfaces

*Masaki Tanaka*, Rena Sugimoto, Nobuhumi Nakamura*

Department of Biotechnology and Life Science, Faculty of engineering, Tokyo University of Agriculture and Technology, 2-24-16, Naka-cho, Koganei, Tokyo, 184-8588, Japan

E-mail: m-tanaka@me.tuat.ac.jp

**Supporting information**

**Experimental details**

**Materials and general methods.** All reagents were used as purchased from commercial sources without further purification. All compounds were purified by column chromatography followed by temperature-gradient vacuum sublimation. Nuclear magnetic resonance (NMR) spectra were obtained in $CDCl_3$ using a JNM-ECX400 NMR spectrometer (JEOL) at ambient temperature. Absorption spectra of organic films on a quartz glass substrate were measured on UV-2550 (Shimadzu). A glass transition temperature was determined using differential scanning calorimetry on DSC7000X (Hitachi).

**Film sample fabrication and evaluation.** Organic films of varying thicknesses for surface potential measurements were deposited directly on pre-cleaned 100-nm-thick ITO-coated glass substrates using physical vapor deposition. Vacuum deposition was performed under high vacuum at pressure levels $< 3.0 \times 10^{-4}$ Pa at a monitored deposition rate using an in-house evaporation machine directly connected to the chamber mounted with a Kelvin probe. The deposition rate was basically controlled at 0.1 nm $s^{-1}$. To calculate the orientation degrees of polar molecules in codeposited films, the film densities of mixed films were assumed to be the weighted average densities of polar and host molecules. The surface potential was measured using the Kelvin probe method under vacuum and dark conditions (UHVKP020, K.P. Technology). For the basic measurement of the surface potentials to estimate the GSP slopes, the samples were transferred between the vacuum chamber and the measurement chamber without air exposure, and the film thickness was increased by vacuum deposition on top of the sample. Another vacuum chamber with a $T_s$ control system was used to test the dependence of the GSP slope on $T_s$. After the film was prepared on an ITO substrate, the film samples were carried out of the deposition chamber to ambient air and immediately transferred to the chamber with a Kelvin probe to measure the surface potential. For this experiment, we prepared film samples of different thicknesses to estimate the GSP slope. The thickness of the deposited film was estimated using a thickness meter (FR-ES, ThetaMetrisis). The HOMO levels were estimated using a photoelectron yield spectrometer (PYS). The PYS measurements of 100-nm-thick films on ITO-coated substrates were performed using BIP-KV100 (Bunkoukeiki).

**Device fabrication and characterization.** HODs were fabricated by the vacuum deposition process without exposure to ambient air. All organic layers were deposited at a deposition rate of 0.1 nm $s^{-1}$. The deposition rates of $MoO_x$ and Al layers was 0.2 and 0.5 nm $s^{-1}$. $MoO_3$ was deposited as $MoO_x$ layers. All device characterizations were performed at room temperature. Current density-voltage measurements were performed using a sourcemeter (Keithley 2400). For photocurrent measurements, UV light was irradiated through the ITO electrode side of HODs using a 365 nm LED light source (M365L3, thorlabs) with a long-pass filter.

**Computational calculations.** Optimized molecular structures and permanent dipoles of ground-state molecules were calculated using the B3LYP/6-31 G (d) level with the Gaussian 16 program package. Conformation analysis was performed using CONFLEX 9 with MMFF94s force field. To calculate atomic polarizabilities, structural optimization was initially conducted using the B3LYP/6-311G(3d, 3p) level. Subsequent calculations were performed to generate the wavefunctions of the molecules under both zero and static fields. Utilizing the wavefunction files, the atomic dipole moments in the presence of these fields were determined using AIMAll software. The polarizabilities were then derived based on the variation in the induced dipole moments with the applied fields, and the isotropic polarizability ($a_{iso}$) was calculated as the average of the diagonal elements of the tensors.

**Synthesis**

FDI-2TFB: 9,9-Bis(3,4-dicarboxyphenyl)fluorene dianhydride (768 mg), 4'-(trifluoromethyl)-biphenyl-4-amine (776 mg), benzoic acid (408 mg), and molecular sieve were added to 1,3-dimethyl-2-imidazolidinone (8 mL). The solution was stirred at room temperature for 2 h and subsequently stirred at 150 °C for 20 h. The resulting solution was washed with water. The precipitate was dried under vacuum, and purified by chromatography on silica gel (chloroform:hexane:ethyl acetate = 18:1:1) to afford FDI-2TFB as a white solid in 64% yield. $^1$H NMR (400 MHz, $CDCl_3$): $\delta$ 7.86 (t, $J$ = 7.5 Hz, 6H), 7.71 (m, 12H), 7.78 (dd, $J$ = 8.2, 1.6 Hz, 2H), 7.51 (m, 6H) 7.39 (m, 4H). $^{13}$C NMR (100 MHz, $CDCl_3$): $\delta$ 167.03, 166.73, 152.96, 148.41, 143.79, 140.38, 139.70, 133.79, 132.59, 131.64, 130.67, 129.97, 129.64, 129.09, 128.86, 128.17, 127.59, 127.06, 125.96, 125.96, 125.92, 125.87, 125.67, 124.42, 123.46, 123.47, 122.96, 121.22, 66.13.

FDI-2bTFB: 9,9-Bis(3,4-dicarboxyphenyl)fluorene dianhydride (768 mg), 4-amino-3',5'-bis(trifluoromethyl)biphenyl (1024 mg), benzoic acid (480 mg), and molecular sieve were added to 1,3-dimethyl-2-imidazolidinone (6 mL). The solution was stirred at room temperature for 20 min and subsequently stirred at 150 °C for 10 h. The resulting solution was washed with water. The precipitate was dried under vacuum, and purified by chromatography on silica gel (chloroform) to afford FDI-2bTFB as a white solid in 66% yield. $^1$H NMR (400 MHz, $CDCl_3$): $\delta$ 8.03 (s, 4H), 7.80 (m, 8H), 7.78 (d, $J$ = 8.7 Hz, 4H), 7.60 (m, 6H) 7.50 (t, $J$ = 6.4 Hz, 2H), 7.39 (m, 4H). $^{13}$C NMR (100 MHz, $CDCl_3$): $\delta$ 166.89, 166.59, 153.02, 148.35, 142.40, 140.37, 138.03, 133.84, 132.53, 132.21, 130.63, 129.11, 128.86, 128.16, 127.27, 125.83, 124.72, 124.47, 123.50, 122.01, 121.43, 121.23, 66.11.

FDI-2FC$_3$: 9,9-Bis(3,4-dicarboxyphenyl)fluorene dianhydride (570 mg), 1*H*,1*H*-heptafluorobutylamine (498 mg), benzoic acid (305 mg), and molecular sieve were added to 1,3-dimethyl-2-imidazolidinone (5 mL). The solution was stirred at room temperature for 5 h and subsequently stirred at 70 °C for 20 h. The resulting solution was washed with water. The precipitate was dried under vacuum, and purified by chromatography on silica gel (chloroform) to afford FDI-2FC$_3$ as a white solid in 71% yield. $^1$H NMR (400 MHz, $CDCl_3$): $\delta$ 7.94 (dd, $J$ = 10, 7.8 Hz, 4H), 7.73 (d, $J$ =1.8 Hz, 2H), 7.59 (dd, $J$ = 7.8, 1.8 Hz, 2H), 7.46 (m, 2H), 7.34 (m, 4H), 4.34 (t, $J$ = 15 Hz, 4H). $^{13}$C NMR (100 MHz, $CDCl_3$): $\delta$ 166.61, 166.29, 152.96, 148.13, 140.32, 133.94, 132.40, 130.54, 129.13, 128.86, 125.78, 124.48, 123.38, 121.20, 65.99, 37.46, 37.21, 36.98.

FDI-2FC$_5$: 9,9-Bis(3,4-dicarboxyphenyl)fluorene dianhydride (570 mg), 1*H*,1*H*-undecafluorohexylamine (837 mg), benzoic acid (308 mg), and molecular sieve were added to 1,3-dimethyl-2-imidazolidinone (5 mL). The solution was stirred at room temperature for 5 h and subsequently stirred at 70 °C for 20 h. The resulting solution was washed with water. The precipitate was dried under vacuum, and purified by chromatography on silica gel (chloroform) to afford FDI-2FC$_5$ as a white solid in 77% yield. $^1$H NMR (400 MHz, $CDCl_3$): $\delta$ 7.83 (dd, $J$ = 10, 7.8 Hz, 4H), 7.73 (s, 2H), 7.59 (dd, $J$ = 7.8, 1.8 Hz, 2H), 7.47 (m, 2H), 7.35 (m, 4H), 4.34 (t, $J$ = 15 Hz, 4H). $^{13}$C NMR (100 MHz, $CDCl_3$): $\delta$ 166.62, 166.29, 152.97, 148.13, 140.32,

133.94, 132.40, 130.54, 129.13, 128.86, 125.78, 124.48, 123.38, 121.20, 65.99, 37.67, 37.43, 37.21.

**Table S1. GSP slope values of reported molecules**

**Table S2.** GSP slope values shown in Figure 1 (b).

| | PDM (Debye) | GSP slope (mV nm$^{-1}$) |
|---|---|---|
| $Alq_3$[1] | 4.4 | +48 |
| *p*-ethyl-TPBi[2] | 7.0 | +141 |
| BCPO[3] | 3.5 | +151 |
| 6F-2TRZ[4] | 2.97 | -108 |
| 6F-2Cz[4] | 0.98 | −40 |
| 6F-TPA-TRZ[4] | 3.19 | −46 |
| 6F-Cz-TRZ[4] | 2.53 | −63 |
| 6F-2BN[4] | 3.16 | +69 |
| 3F-3BN[4] | 3.02 | +130 |
| 6F-Cz-TRZ[4] | 2.53 | −61 |
| 6FDI-2BTA[5] | 2.26 | +66.0 |
| 6FDI-2TAZ[5] | 3.65 | +152 |
| 6FDI-2*p*BN[5] | 5.84 | +218 |
| 6FDI-2*p*BNMe[5] | 5.55 | +217 |
| 6FDI-2*o*BN[5] | 4.61 | −103 |
| 6FDI-2*m*BN[5] | 5.93 | +87 |
| 6F-2PI[5] | 5.82 | −210 |
| FL-2PI[6] | 2.53 | −31.8 |
| 6H-2PI[6] | 2.58 | −59.4 |
| CH-2PI[6] | 2.70 | −88.0 |
| SO-2PI[6] | 9.35 | +25.3 |
| 6F-2PIMe[6] | 6.49 | −124 |
| 6F-2PItBu[6] | 6.83 | +21.7 |
| SODI-2CH[6] | 3.66 | +162 |
| SO-2PItBu[6] | 10.5 | +321 |

**Figure S1–8. NMR results**

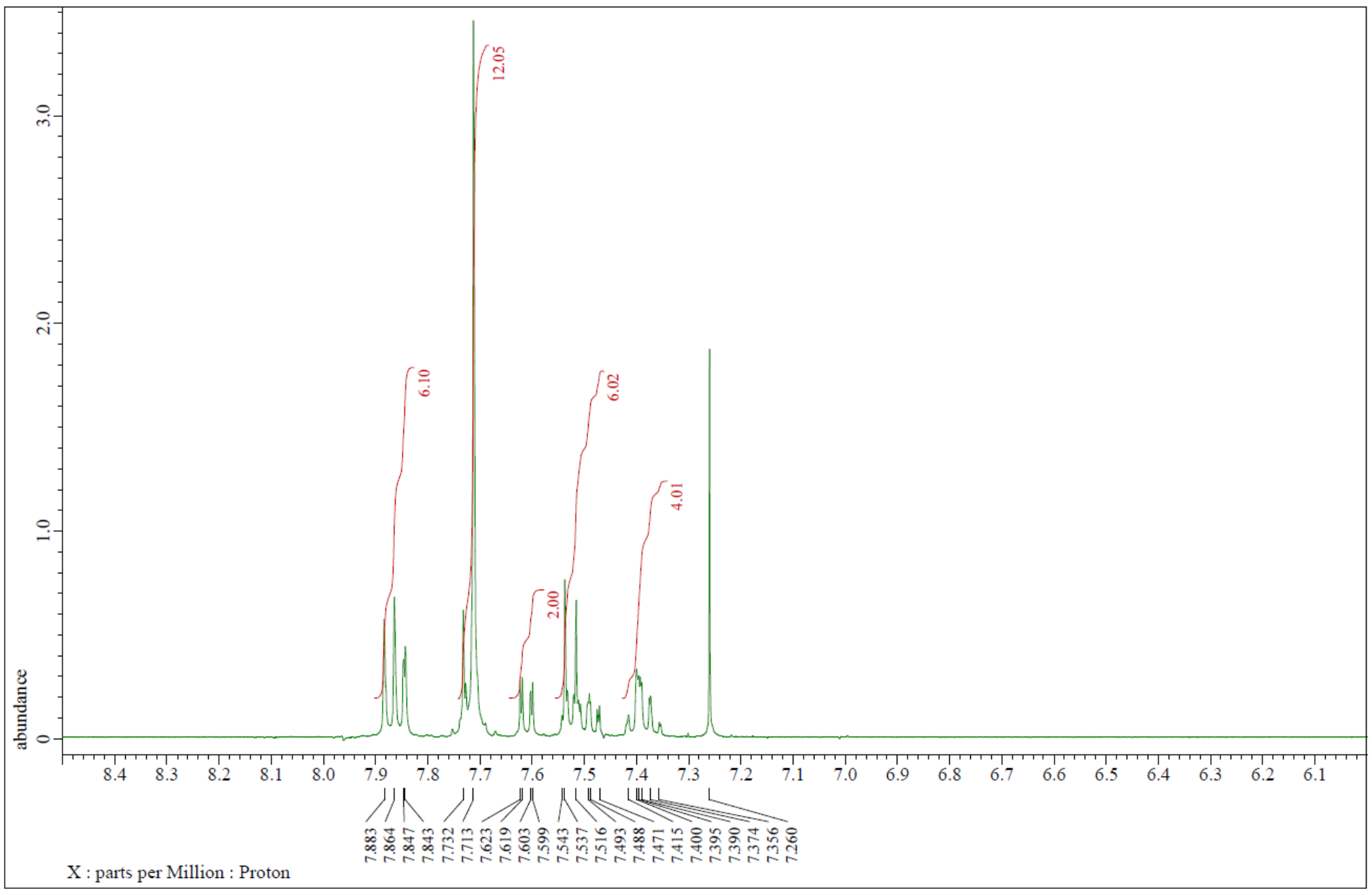

**Figure S7.** $^{1}H$ NMR spectrum of FDI-2TFB.

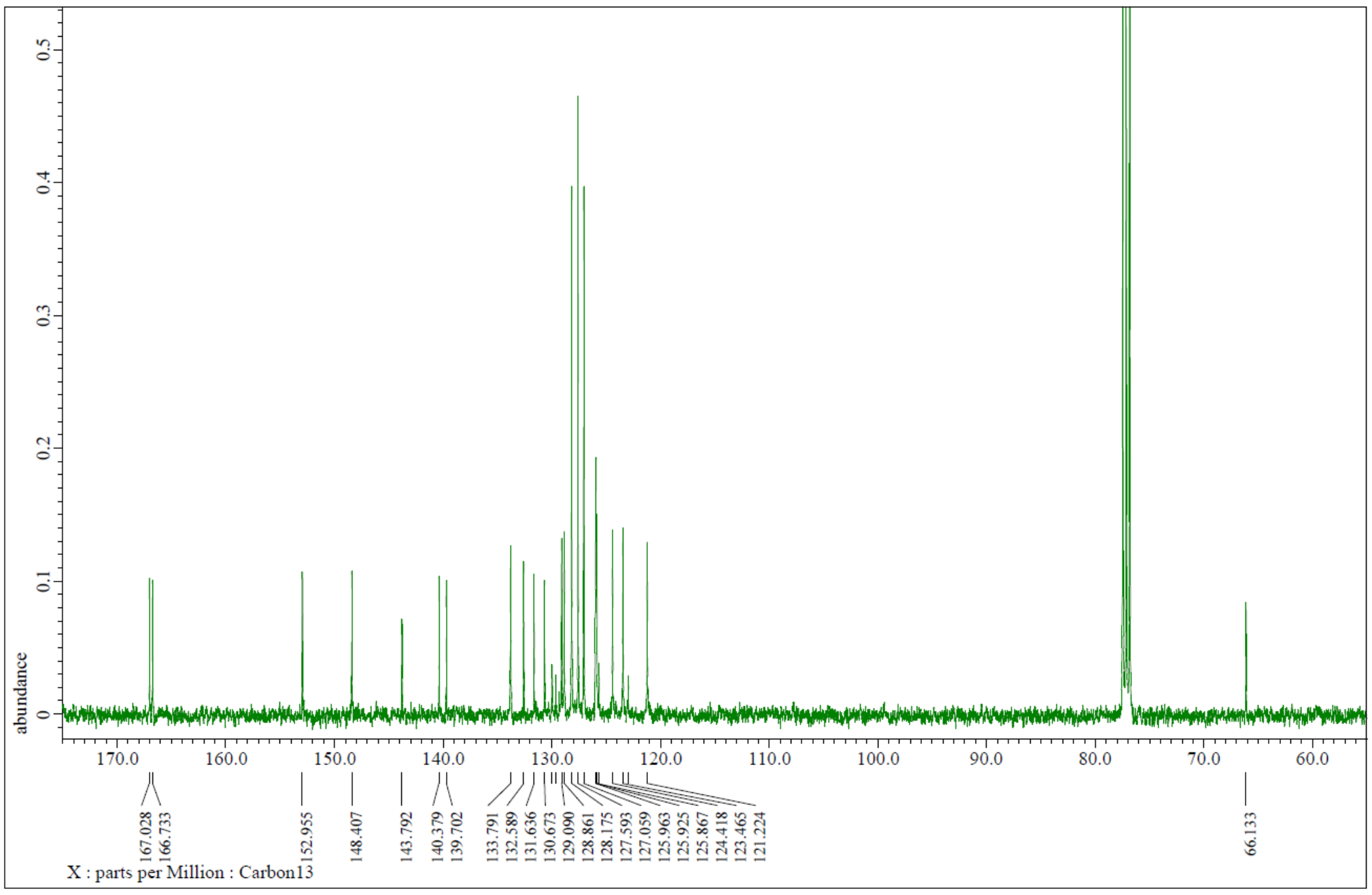

**Figure S8.** $^{13}C$ NMR spectrum of FDI-2TFB.

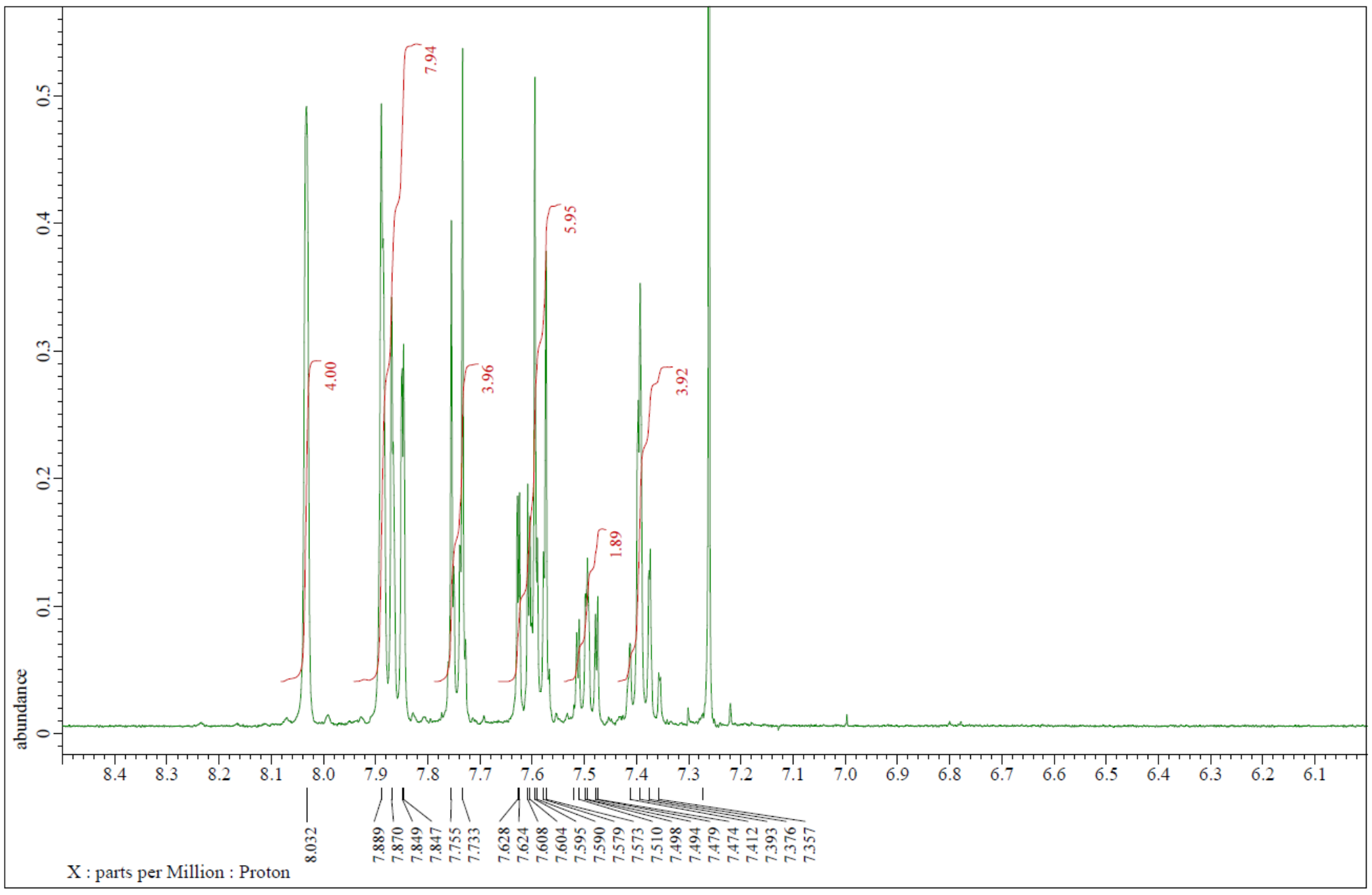

**Figure S9.** $^{1}H$ NMR spectrum of FDI-2bTFB.

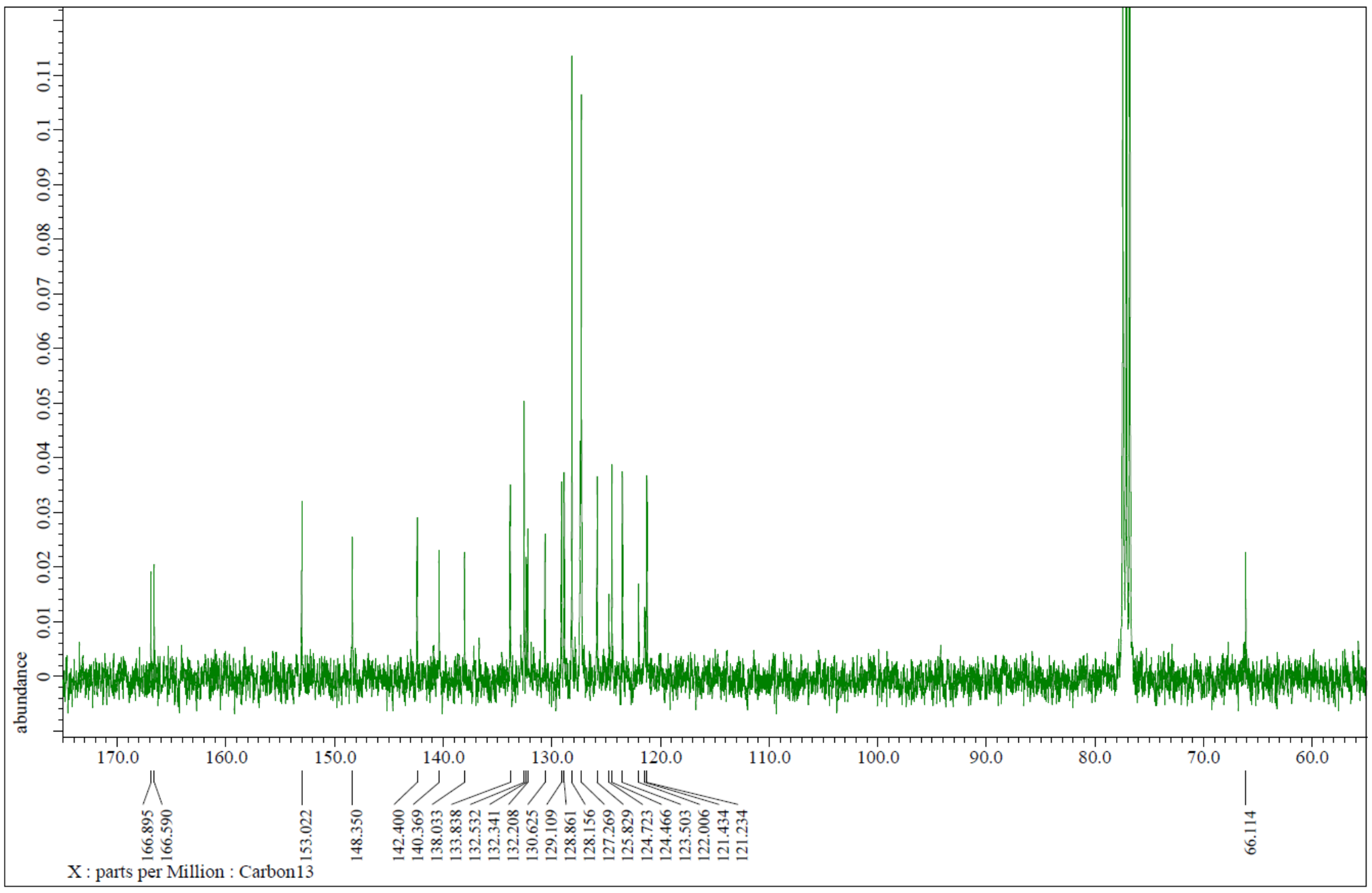

**Figure S10.** $^{13}C$ NMR spectrum of FDI-2bTFB.

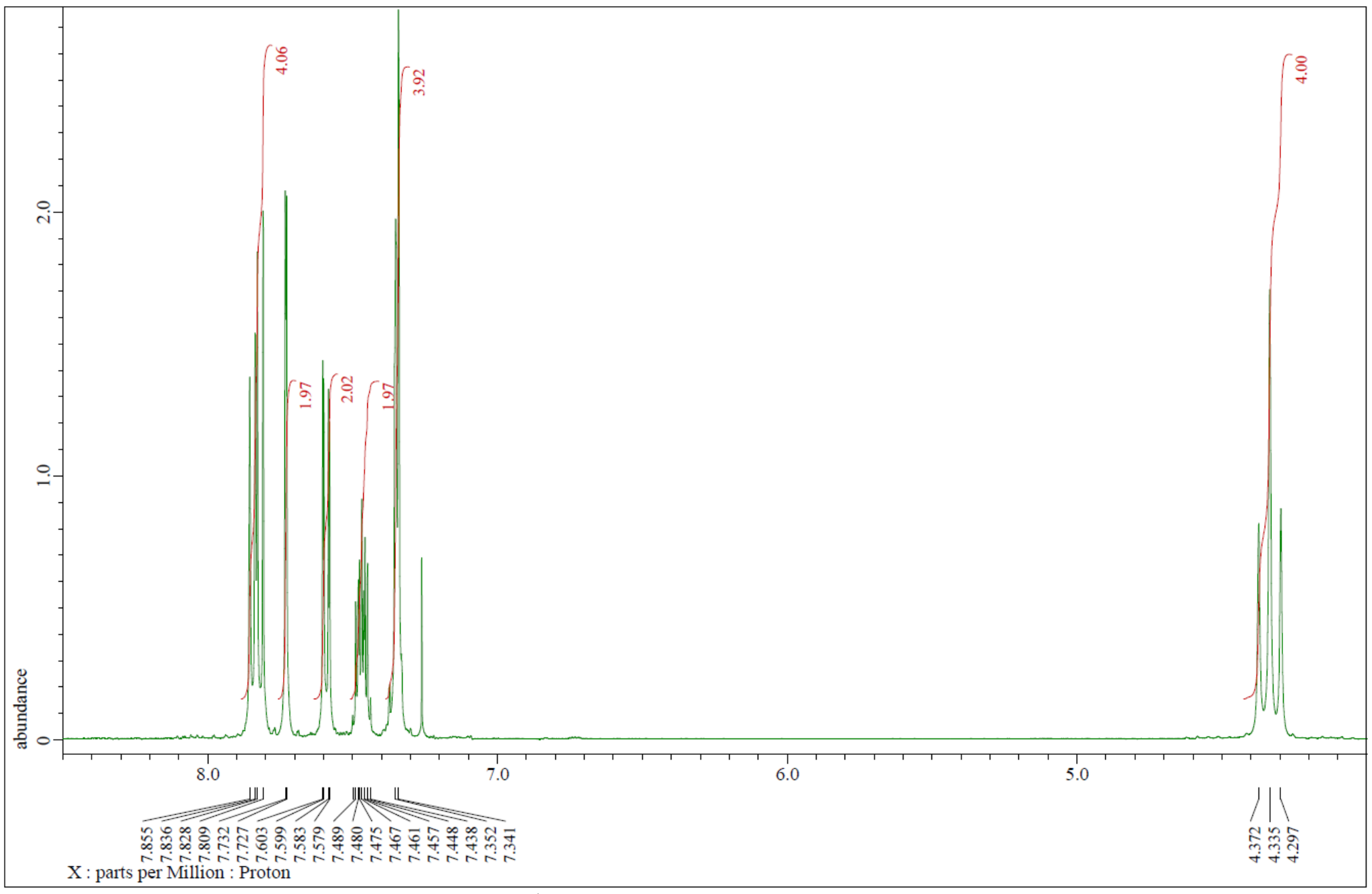

**Figure S11.** $^{1}H$ NMR spectrum of FDI-2FC$_3$.

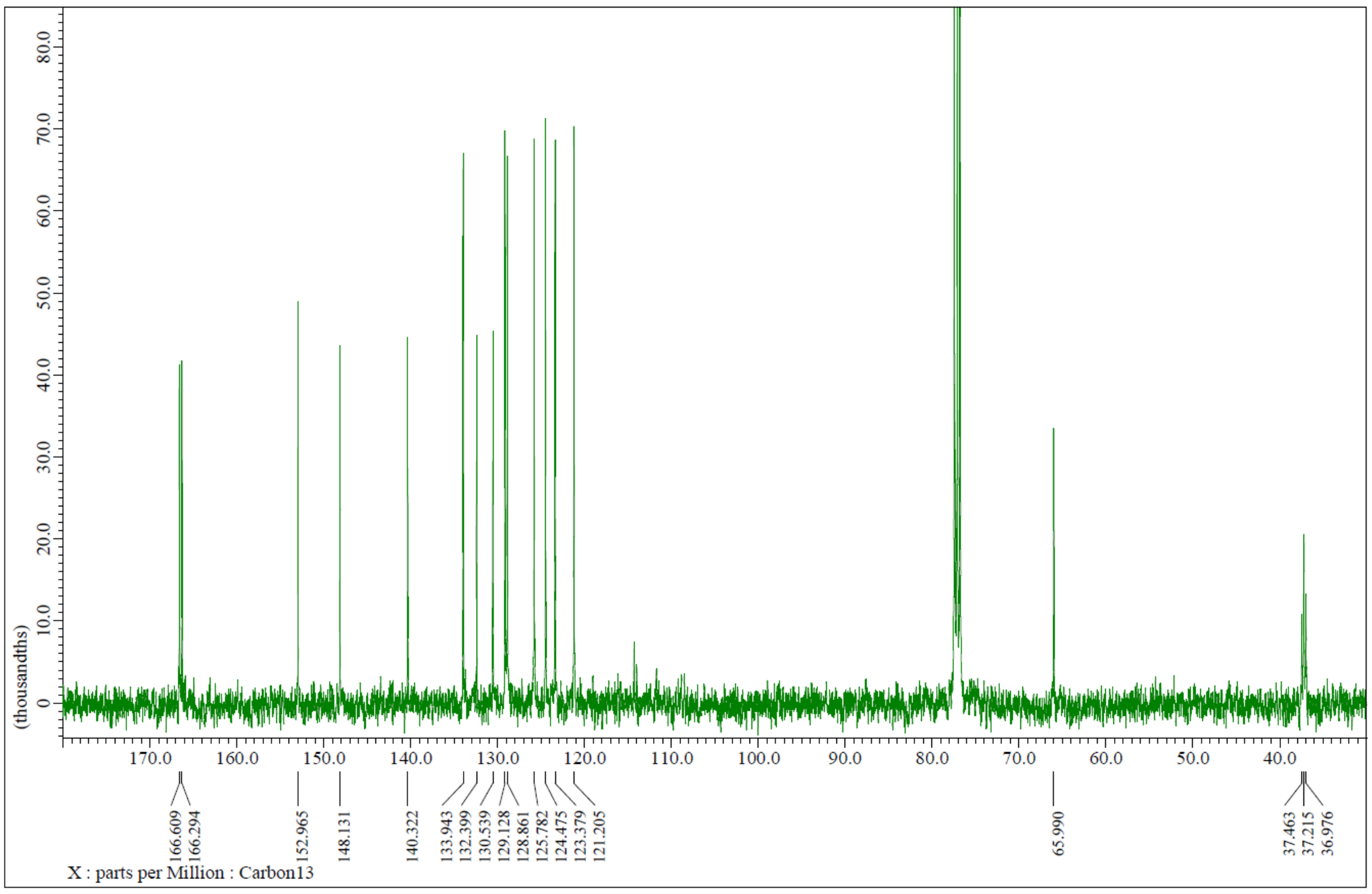

**Figure S12.** $^{13}C$ NMR spectrum of FDI-2FC$_3$.

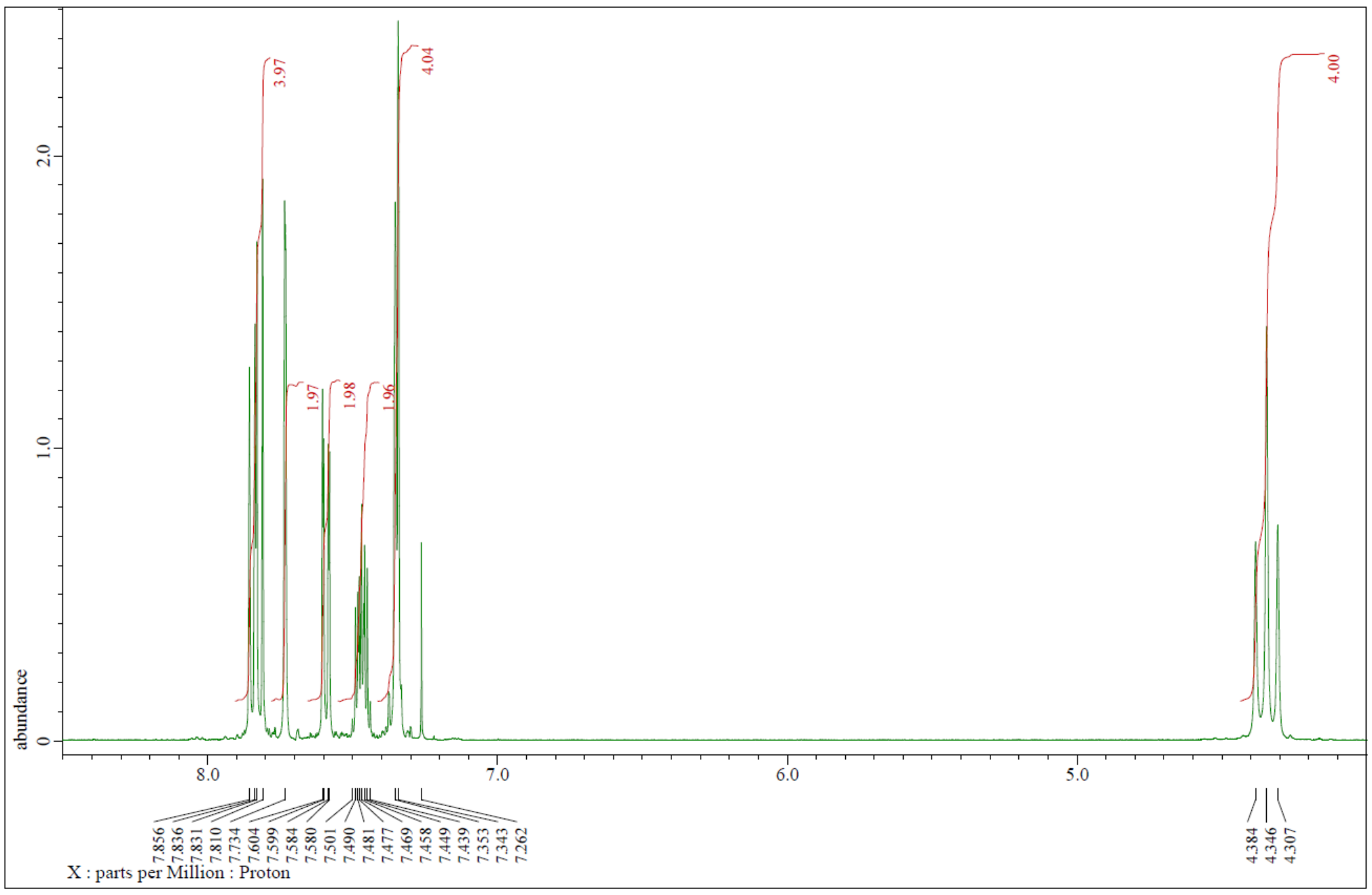

**Figure S13.** $^{1}$H NMR spectrum of FDI-2FC$_5$.

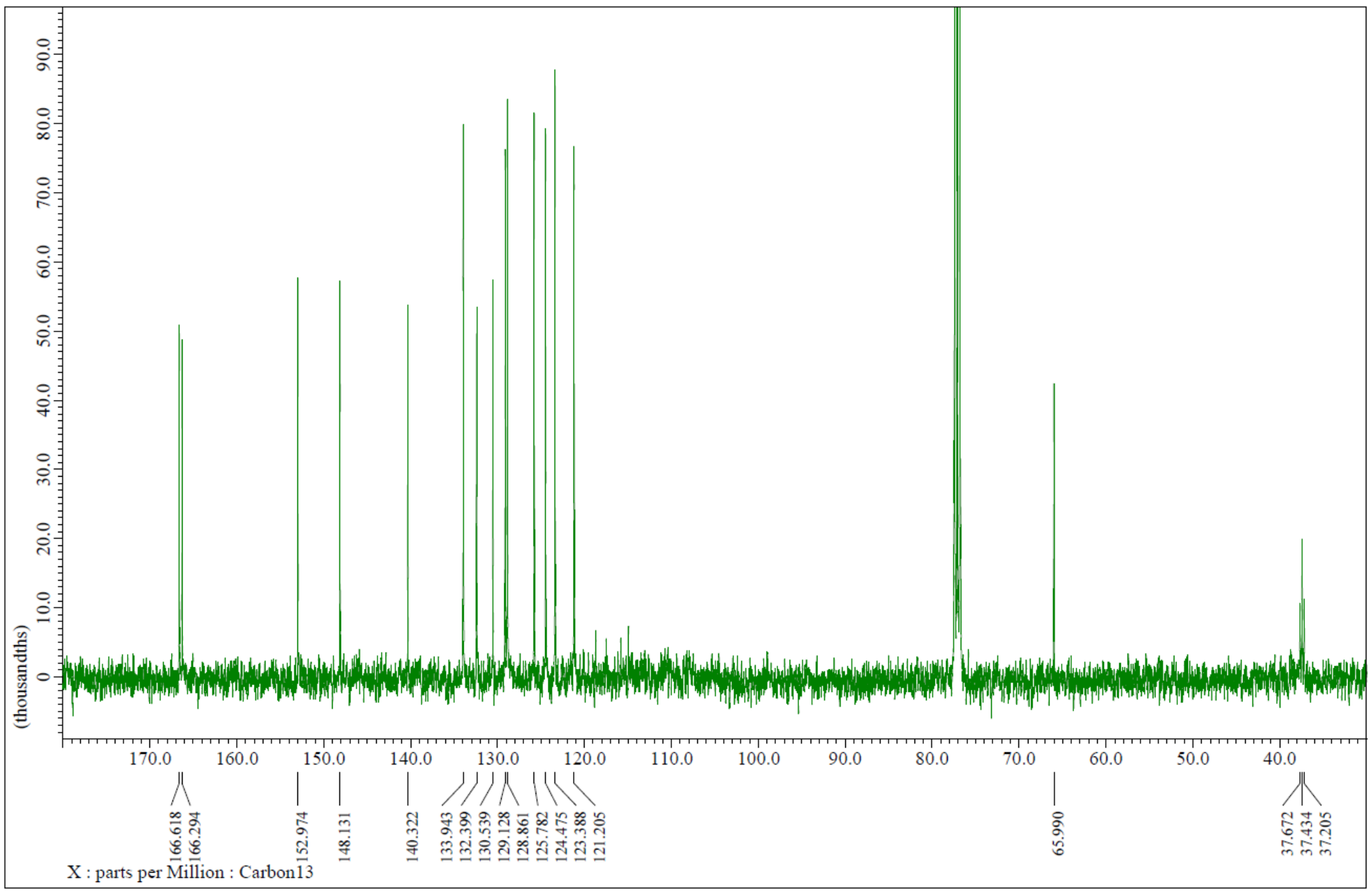

**Figure S14.** $^{13}$C NMR spectrum of FDI-2FC$_5$.

**Figure S9. Dipole moment direction**

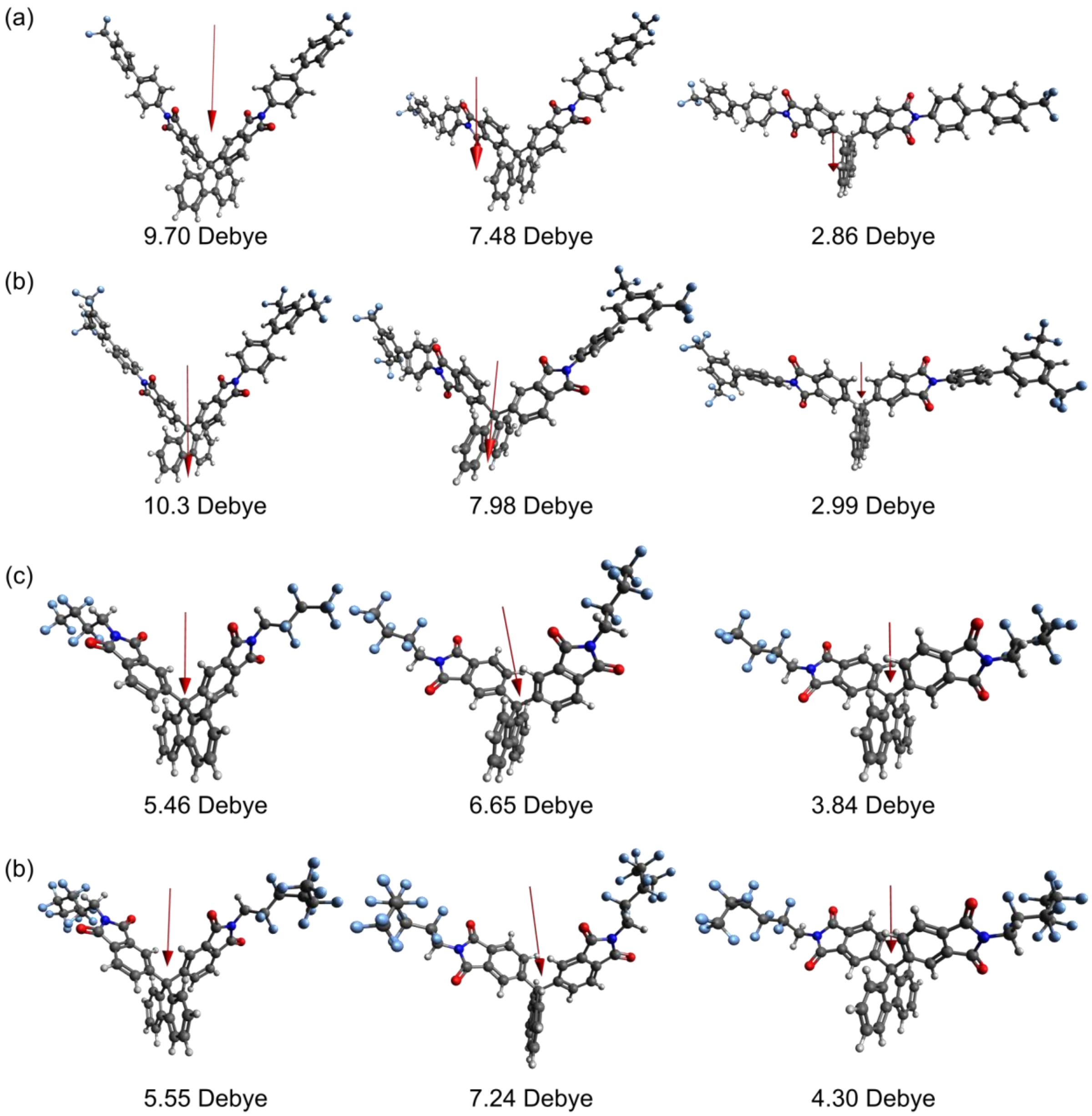

**Figure S15.** Permanent dipole moment direction. (a) FDI-2TFB. (b) FDI-2bTFB. (c) FDI-2FC$_3$. (d) FDI-2FC$_5$.

**Figure S10. Distribution of dipole magnitude**

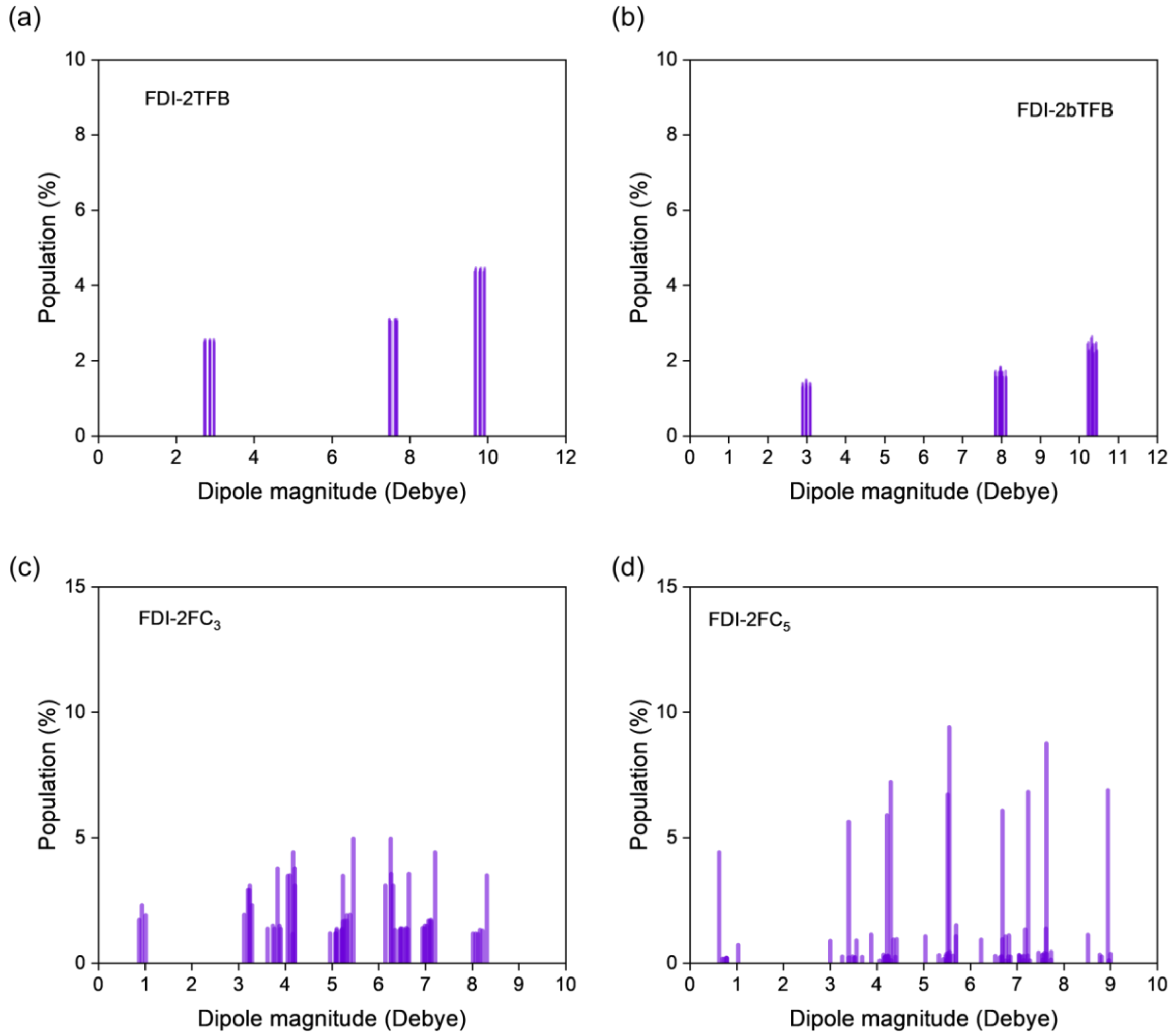

**Figure S16.** Dipole magnitude distribution. (a) FDI-2TFB. (b) FDI-2bTFB. (c) FDI-2FC$_3$. (d) FDI-2FC$_5$.

**Figure S11. Atomic polarizability calculation of FDI-2TFB**

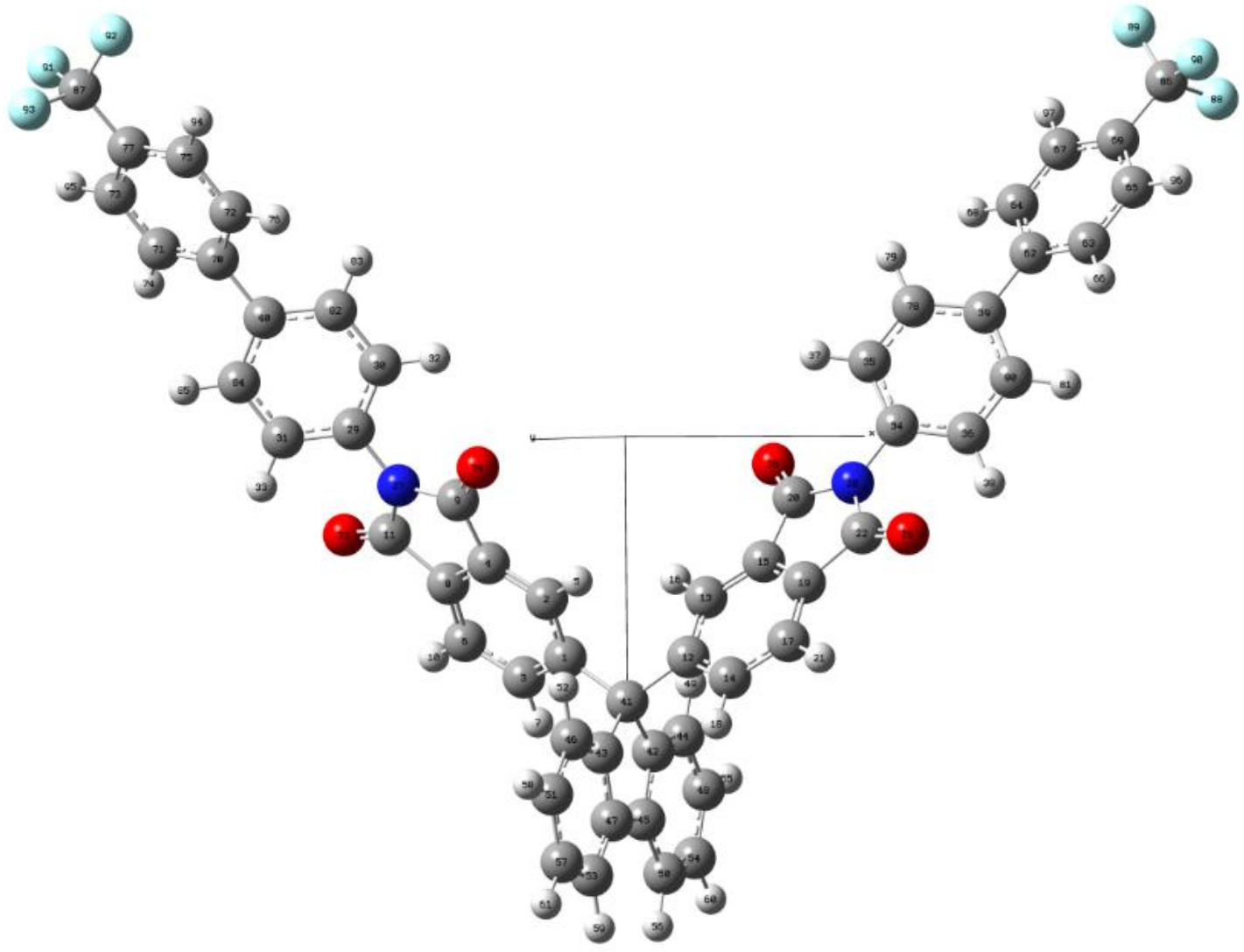

**Figure S17.** Molecular model of FDI-2TFB for calculation of atomic polarizabilities.

**Table S2. Atomic polarizabilities of FDI-2TFB**

**Table S3.** Calucurated atomic polarizabilities of FDI-2TFB.

| | $a_{\mathrm{iso}}$ ($10^{-40}$ C m$^2$ V$^{-1}$) | | $a_{\mathrm{iso}}$ ($10^{-40}$ C m$^2$ V$^{-1}$) | | $a_{\mathrm{iso}}$ ($10^{-40}$ C m$^2$ V$^{-1}$) | | $a_{\mathrm{iso}}$ ($10^{-40}$ C m$^2$ V$^{-1}$) |
|---|---|---|---|---|---|---|---|
| C1 | 2.0680 | C31 | 1.7542 | H61 | 0.4874 | F91 | 0.6883 |
| C2 | 1.8403 | H32 | 0.4640 | C62 | 2.2182 | F92 | 0.6798 |
| C3 | 1.6230 | H33 | 0.4611 | C63 | 1.6806 | F93 | 0.6942 |
| C4 | 2.0228 | C34 | 1.6697 | C64 | 1.6491 | H94 | 0.3870 |
| H5 | 0.3119 | C35 | 1.7088 | C65 | 1.5895 | H95 | 0.3983 |
| C6 | 1.7572 | C36 | 1.7532 | H66 | 0.3549 | H96 | 0.3983 |
| H7 | 0.3492 | H37 | 0.4636 | C67 | 1.5933 | H97 | 0.3870 |
| C8 | 1.8925 | H38 | 0.4591 | H68 | 0.3517 | | |
| C9 | 1.2457 | C39 | 2.2350 | C69 | 1.6168 | | |
| H10 | 0.3779 | C40 | 2.2399 | C70 | 2.2158 | | |
| C11 | 1.2699 | C41 | 2.7493 | C71 | 1.6820 | | |
| C12 | 2.0665 | C42 | 1.8368 | C72 | 1.6473 | | |
| C13 | 1.8452 | C43 | 1.8357 | C73 | 1.5897 | | |
| C14 | 1.6188 | C44 | 1.5421 | H74 | 0.3547 | | |
| C15 | 2.0241 | C45 | 1.8783 | C75 | 1.5913 | | |
| H16 | 0.3119 | C46 | 1.5430 | H76 | 0.3517 | | |
| C17 | 1.7564 | C47 | 1.8796 | C77 | 1.6142 | | |
| H18 | 0.3494 | C48 | 1.5347 | C78 | 1.7857 | | |
| C19 | 1.8928 | H49 | 0.3547 | H79 | 0.3650 | | |
| C20 | 1.2535 | C50 | 1.6008 | C80 | 1.7612 | | |
| H21 | 0.3779 | C51 | 1.5338 | H81 | 0.3839 | | |
| C22 | 1.2691 | H52 | 0.3549 | C82 | 1.7842 | | |
| O23 | 1.3133 | C53 | 1.6004 | H83 | 0.3650 | | |
| O24 | 1.2098 | C54 | 1.4560 | C84 | 1.7620 | | |
| O25 | 1.2091 | H55 | 0.4443 | H85 | 0.3839 | | |
| O26 | 1.3129 | H56 | 0.4152 | C86 | 0.8050 | | |
| N27 | 2.6172 | C57 | 1.4558 | C87 | 0.8067 | | |
| N28 | 2.6124 | H58 | 0.4443 | F88 | 0.6942 | | |
| C29 | 1.6660 | H59 | 0.4152 | F89 | 0.6798 | | |
| C30 | 1.7083 | H60 | 0.4874 | F90 | 0.6883 | | |

**Figure S12. Atomic polarizability calculation of FDI-2bTFB**

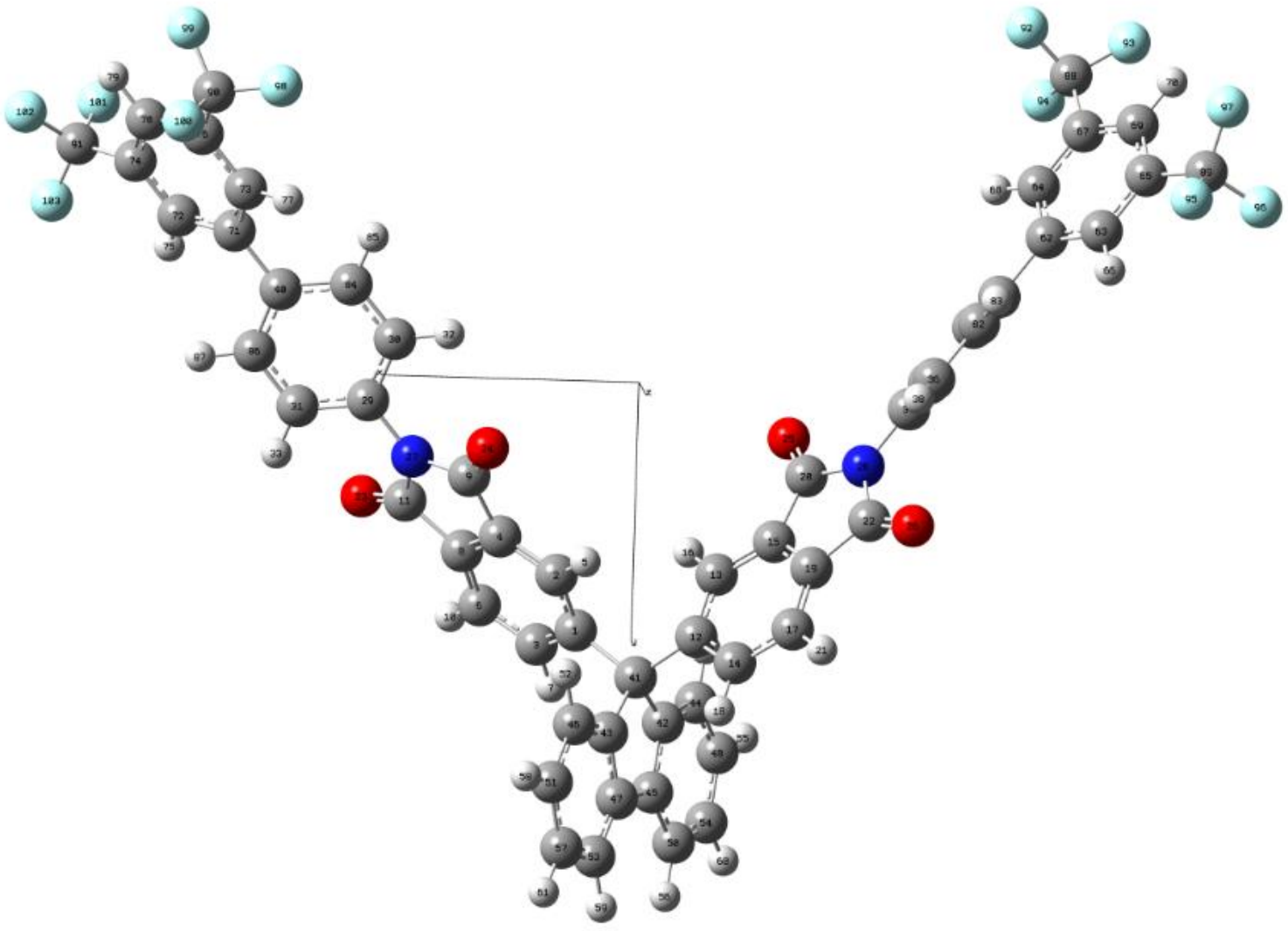

**Figure S18.** Molecular model of FDI-2bTFB for calculation of atomic polarizabilities.

**Table S3. Atomic polarizabilities of FDI-2bTFB**

**Table S4.** Calucurated atomic polarizabilities of FDI-2bTFB.

| | $a_{iso}$ ($10^{-40}$ C m$^2$ V$^{-1}$) | | $a_{iso}$ ($10^{-40}$ C m$^2$ V$^{-1}$) | | $a_{iso}$ ($10^{-40}$ C m$^2$ V$^{-1}$) | | $a_{iso}$ ($10^{-40}$ C m$^2$ V$^{-1}$) |
|---|---|---|---|---|---|---|---|
| C1 | 2.0703 | C31 | 1.7541 | H61 | 0.4868 | C91 | 0.7135 |
| C2 | 1.8444 | H32 | 0.4592 | C62 | 2.2541 | F92 | 0.6486 |
| C3 | 1.6236 | H33 | 0.4569 | C63 | 1.6642 | F93 | 0.6872 |
| C4 | 2.0255 | C34 | 1.6632 | C64 | 1.6234 | F94 | 0.6232 |
| H5 | 0.3106 | C35 | 1.7173 | C65 | 1.7008 | F95 | 0.6299 |
| C6 | 1.7556 | C36 | 1.7393 | H66 | 0.3139 | F96 | 0.6602 |
| H7 | 0.3491 | H37 | 0.4600 | C67 | 1.6948 | F97 | 0.6880 |
| C8 | 1.8914 | H38 | 0.4525 | H68 | 0.3123 | F98 | 0.6220 |
| C9 | 1.2450 | C39 | 2.2158 | C69 | 1.4641 | F99 | 0.6838 |
| H10 | 0.3764 | C40 | 2.2140 | H70 | 0.3848 | F100 | 0.6521 |
| C11 | 1.2709 | C41 | 2.7576 | C71 | 2.2501 | F101 | 0.6559 |
| C12 | 2.0776 | C42 | 1.8307 | C72 | 1.6562 | F102 | 0.6904 |
| C13 | 1.8508 | C43 | 1.8347 | C73 | 1.6269 | F103 | 0.6305 |
| C14 | 1.6241 | C44 | 1.5383 | C74 | 1.6983 | | |
| C15 | 2.0326 | C45 | 1.8750 | H75 | 0.3135 | | |
| H16 | 0.3112 | C46 | 1.5418 | C76 | 1.6976 | | |
| C17 | 1.7622 | C47 | 1.8751 | H77 | 0.3118 | | |
| H18 | 0.3488 | C48 | 1.5310 | C78 | 1.4657 | | |
| C19 | 1.8971 | H49 | 0.3557 | H79 | 0.3844 | | |
| C20 | 1.2595 | C50 | 1.5961 | C80 | 1.7705 | | |
| H21 | 0.3766 | C51 | 1.5302 | H81 | 0.3677 | | |
| C22 | 1.2722 | H52 | 0.3549 | C82 | 1.7597 | | |
| O23 | 1.3120 | C53 | 1.5947 | H83 | 0.3744 | | |
| O24 | 1.2060 | C54 | 1.4527 | C84 | 1.7772 | | |
| O25 | 1.2144 | H55 | 0.4431 | H85 | 0.3618 | | |
| O26 | 1.3104 | H56 | 0.4149 | C86 | 1.7530 | | |
| N27 | 2.6183 | C57 | 1.4517 | H87 | 0.3812 | | |
| N28 | 2.6197 | H58 | 0.4431 | C88 | 0.7081 | | |
| C29 | 1.6650 | H59 | 0.4150 | C89 | 0.7176 | | |
| C30 | 1.7052 | H60 | 0.4869 | C90 | 0.7017 | | |

**Figure S13. Atomic polarizability calculation of FDI-2FC$_3$**

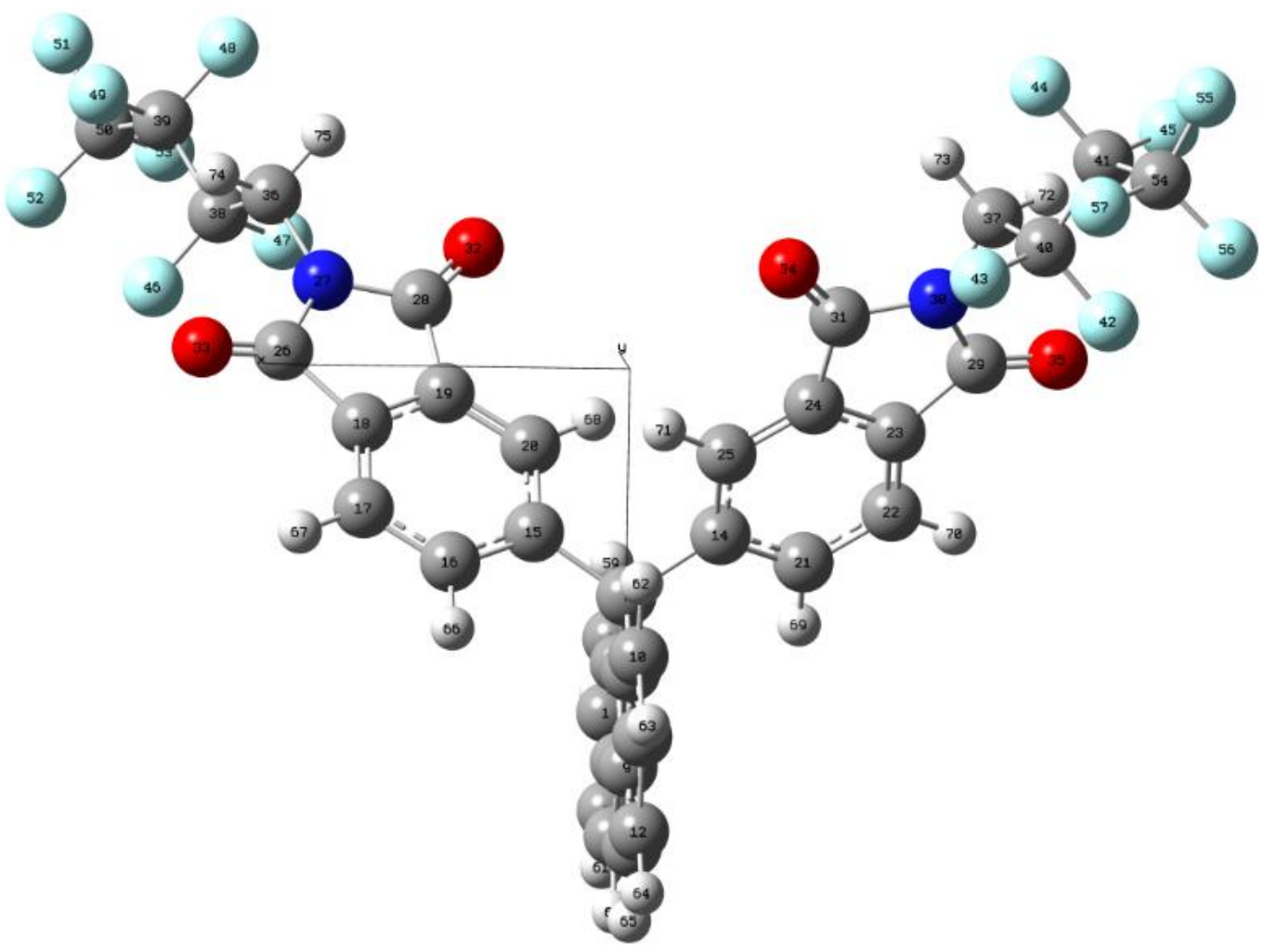

**Figure S19.** Molecular model of FDI-2FC$_3$ for calculation of atomic polarizabilities.

**Table S4. Atomic polarizabilities of FDI-2FC3**

**Table S5.** Calucurated atomic polarizabilities of FDI-2FC3.

| | $a_{iso}$ ($10^{-40}$ C m$^2$ V$^{-1}$) | | $a_{iso}$ ($10^{-40}$ C m$^2$ V$^{-1}$) | | $a_{iso}$ ($10^{-40}$ C m$^2$ V$^{-1}$) |
|---|---|---|---|---|---|
| C1 | 1.5064 | C31 | 1.0905 | H61 | 0.4748 |
| C2 | 1.7699 | O32 | 1.0887 | H62 | 0.3463 |
| C3 | 1.8210 | O33 | 1.1960 | H63 | 0.4435 |
| C4 | 1.5677 | O34 | 1.0887 | H64 | 0.4747 |
| C5 | 1.4277 | O35 | 1.1960 | H65 | 0.4016 |
| C6 | 2.5500 | C36 | 1.0000 | H66 | 0.3262 |
| C7 | 1.7627 | C37 | 1.0008 | H67 | 0.3726 |
| C8 | 1.8196 | C38 | 0.8764 | H68 | 0.3030 |
| C9 | 1.5053 | C39 | 0.7831 | H69 | 0.3263 |
| C10 | 1.4920 | C40 | 0.8870 | H70 | 0.3726 |
| C11 | 1.4269 | C41 | 0.7869 | H71 | 0.3029 |
| C12 | 1.5639 | F42 | 0.6374 | H72 | 0.3037 |
| C13 | 1.9078 | F43 | 0.6265 | H73 | 0.2926 |
| C14 | 1.9154 | F44 | 0.6063 | H74 | 0.3038 |
| C15 | 1.5170 | F45 | 0.6153 | H75 | 0.2926 |
| C16 | 1.6083 | F46 | 0.6372 | | |
| C17 | 1.6800 | F47 | 0.6264 | | |
| C18 | 1.8044 | F48 | 0.6063 | | |
| C19 | 1.6697 | F49 | 0.6153 | | |
| C20 | 1.5166 | C50 | 0.5072 | | |
| C21 | 1.6047 | F51 | 0.6731 | | |
| C22 | 1.6782 | F52 | 0.6601 | | |
| C23 | 1.7989 | F53 | 0.6562 | | |
| C24 | 1.6614 | C54 | 0.5281 | | |
| C25 | 1.1080 | F55 | 0.6730 | | |
| C26 | 2.0569 | F56 | 0.6605 | | |
| N27 | 1.0927 | F57 | 0.6565 | | |
| C28 | 1.1046 | H58 | 0.4435 | | |
| C29 | 2.0528 | H59 | 0.3466 | | |
| N30 | 1.5064 | H60 | 0.4016 | | |

**Figure S14. Atomic polarizability calculation of FDI-2FC$_5$**

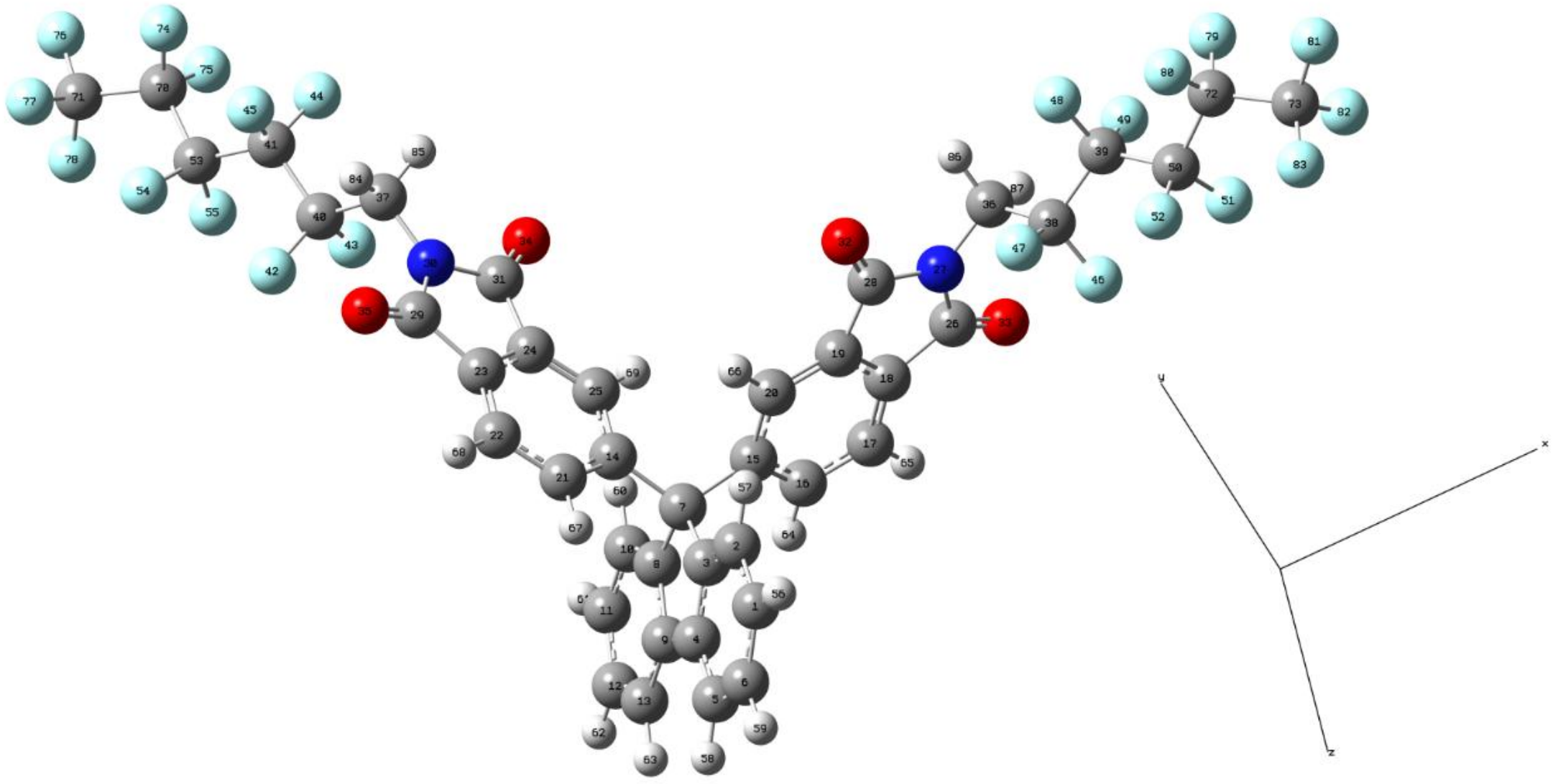

**Figure S20.** Molecular model of FDI-2FC$_5$ for calculation of atomic polarizabilities.

**Table S5. Atomic polarizabilities of FDI-2$FC_5$**

**Table S6.** Calucurated atomic polarizabilities of FDI-2$FC_5$.

| | $a_{iso}$ ($10^{-40}$ C $m^2$ $V^{-1}$) | | $a_{iso}$ ($10^{-40}$ C $m^2$ $V^{-1}$) | | $a_{iso}$ ($10^{-40}$ C $m^2$ $V^{-1}$) |
|---|---|---|---|---|---|
| C1 | 1.5935 | C31 | 1.1234 | H61 | 0.4414 |
| C2 | 1.5516 | O32 | 1.1221 | H62 | 0.4715 |
| C3 | 1.7880 | O33 | 1.2442 | H63 | 0.3917 |
| C4 | 1.8527 | O34 | 1.1224 | H64 | 0.3142 |
| C5 | 1.6433 | O35 | 1.2441 | H65 | 0.3572 |
| C6 | 1.5209 | C36 | 0.9933 | H66 | 0.2860 |
| C7 | 2.6032 | C37 | 0.9908 | H67 | 0.3140 |
| C8 | 1.7894 | C38 | 1.0453 | H68 | 0.3572 |
| C9 | 1.8539 | C39 | 0.9524 | H69 | 0.2859 |
| C10 | 1.5528 | C40 | 1.0428 | C70 | 0.8436 |
| C11 | 1.5955 | C41 | 0.9701 | C71 | 0.6483 |
| C12 | 1.5203 | F42 | 0.5400 | C72 | 0.8408 |
| C13 | 1.6474 | F43 | 0.5966 | C73 | 0.6443 |
| C14 | 1.9303 | F44 | 0.5858 | F74 | 0.6441 |
| C15 | 1.9336 | F45 | 0.6338 | F75 | 0.6023 |
| C16 | 1.5472 | F46 | 0.5403 | F76 | 0.6526 |
| C17 | 1.6625 | F47 | 0.5951 | F77 | 0.6491 |
| C18 | 1.7184 | F48 | 0.6124 | F78 | 0.5787 |
| C19 | 1.8416 | F49 | 0.6329 | F79 | 0.6420 |
| C20 | 1.6784 | C50 | 0.8279 | F80 | 0.6312 |
| C21 | 1.5460 | F51 | 0.6234 | F81 | 0.6526 |
| C22 | 1.6644 | F52 | 0.6242 | F82 | 0.6489 |
| C23 | 1.7165 | C53 | 0.8461 | F83 | 0.5786 |
| C24 | 1.8403 | F54 | 0.6241 | H84 | 0.2796 |
| C25 | 1.6757 | F55 | 0.6244 | H85 | 0.2777 |
| C26 | 1.1489 | H56 | 0.4412 | H86 | 0.2775 |
| N27 | 2.2087 | H57 | 0.3366 | H87 | 0.2794 |
| C28 | 1.1275 | H58 | 0.3918 | | |
| C29 | 1.1499 | H59 | 0.4712 | | |
| N30 | 2.2065 | H60 | 0.3367 | | |

**Figure S15. Atomic polarizability calculation of FDI-2*p*TFP**

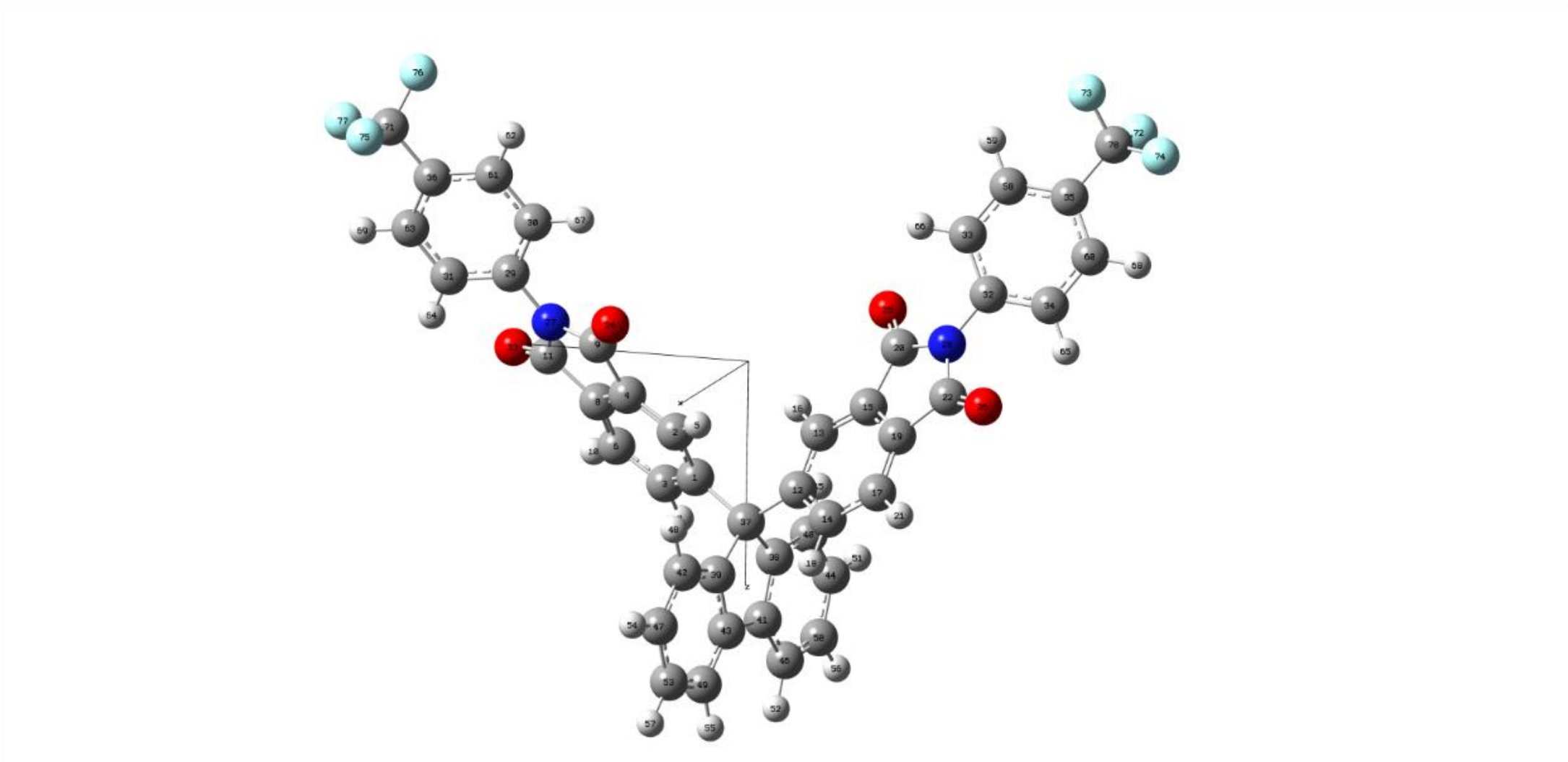

**Figure S21.** Molecular model of FDI-2*p*TFP for calculation of atomic polarizabilities.

**Table S6. Atomic polarizabilities of FDI-2*p*TFP**

**Table S7.** Calucurated atomic polarizabilities of FDI-2*p*TFP.

| | $a_{iso}$ ($10^{-40}$ C m$^2$ V$^{-1}$) | | $a_{iso}$ ($10^{-40}$ C m$^2$ V$^{-1}$) | | $a_{iso}$ ($10^{-40}$ C m$^2$ V$^{-1}$) |
|---|---|---|---|---|---|
| C1 | 2.0023 | C31 | 1.5478 | C61 | 1.5207 |
| C2 | 1.7651 | C32 | 1.5092 | H62 | 0.3723 |
| C3 | 1.5792 | C33 | 1.5100 | C63 | 1.5103 |
| C4 | 1.9343 | C34 | 1.5516 | H64 | 0.4281 |
| H5 | 0.3069 | C35 | 1.5481 | H65 | 0.4285 |
| C6 | 1.6920 | C36 | 1.5458 | H66 | 0.4287 |
| H7 | 0.3381 | C37 | 2.6664 | H67 | 0.4299 |
| C8 | 1.8069 | C38 | 1.8038 | H68 | 0.3904 |
| C9 | 1.1786 | C39 | 1.8058 | H69 | 0.3904 |
| H10 | 0.3741 | C40 | 1.5266 | C70 | 0.7884 |
| C11 | 1.2034 | C41 | 1.8529 | C71 | 0.7863 |
| C12 | 2.0094 | C42 | 1.5264 | F72 | 0.6767 |
| C13 | 1.7739 | C43 | 1.8529 | F73 | 0.6687 |
| C14 | 1.5767 | C44 | 1.5138 | F74 | 0.6824 |
| C15 | 1.9380 | H45 | 0.3501 | F75 | 0.6768 |
| H16 | 0.3068 | C46 | 1.5838 | F76 | 0.6686 |
| C17 | 1.6958 | C47 | 1.5146 | F77 | 0.6824 |
| H18 | 0.3387 | H48 | 0.3502 | | |
| C19 | 1.8108 | C49 | 1.5837 | | |
| C20 | 1.1879 | C50 | 1.4382 | | |
| H21 | 0.3741 | H51 | 0.4434 | | |
| C22 | 1.2043 | H52 | 0.4085 | | |
| O23 | 1.2871 | C53 | 1.4431 | | |
| O24 | 1.1811 | H54 | 0.4434 | | |
| O25 | 1.1817 | H55 | 0.4085 | | |
| O26 | 1.2873 | H56 | 0.4812 | | |
| N27 | 2.3615 | H57 | 0.4812 | | |
| N28 | 2.3668 | C58 | 1.5238 | | |
| C29 | 1.5056 | H59 | 0.3723 | | |
| C30 | 1.5067 | C60 | 1.5142 | | |

**Figure S16. Atomic polarizability calculation of FDI-2*m*TFP**

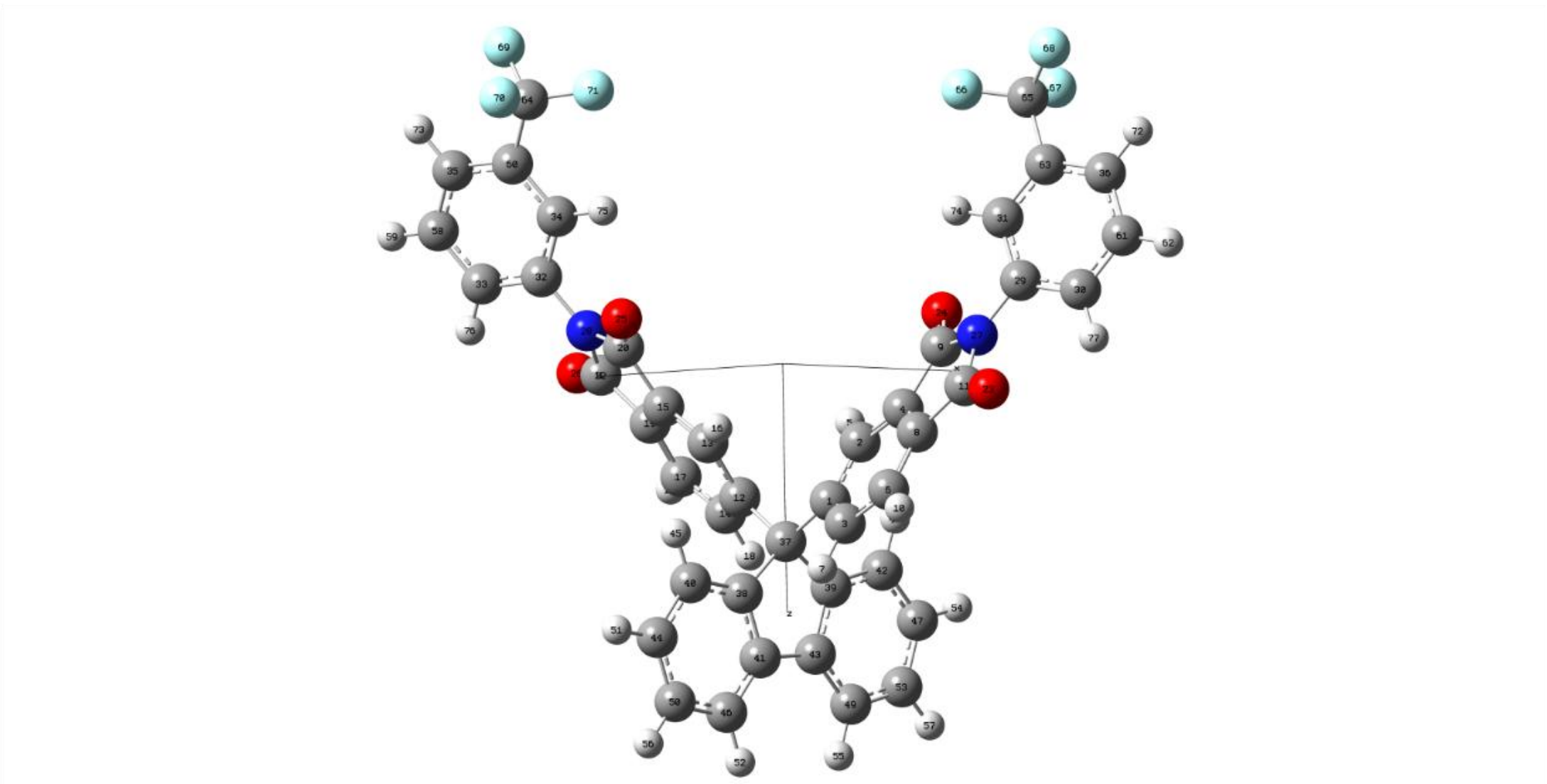

**Figure S22.** Molecular model of FDI-2*m*TFP for calculation of atomic polarizabilities.

**Table S7. Atomic polarizabilities of FDI-2*m*TFP**

**Table S8.** Calucurated atomic polarizabilities of FDI-2*m*TFP.

| | $a_{iso}$ ($10^{-40}$ C m$^2$ V$^{-1}$) | | $a_{iso}$ ($10^{-40}$ C m$^2$ V$^{-1}$) | | $a_{iso}$ ($10^{-40}$ C m$^2$ V$^{-1}$) |
|---|---|---|---|---|---|
| C1 | 2.0273 | C31 | 1.4956 | C61 | 1.5174 |
| C2 | 1.7702 | C32 | 1.5092 | H62 | 0.4411 |
| C3 | 1.5802 | C33 | 1.5079 | C63 | 1.5864 |
| C4 | 1.9337 | C34 | 1.4769 | H64 | 0.6582 |
| H5 | 0.3042 | C35 | 1.3880 | H65 | 0.6746 |
| C6 | 1.6943 | C36 | 1.3974 | H66 | 0.3978 |
| H7 | 0.3375 | C37 | 2.6828 | H67 | 0.6435 |
| C8 | 1.8023 | C38 | 1.7989 | H68 | 0.6791 |
| C9 | 1.1893 | C39 | 1.8035 | H69 | 0.6790 |
| H10 | 0.3747 | C40 | 1.5207 | C70 | 0.6436 |
| C11 | 1.2033 | C41 | 1.8499 | C71 | 0.3608 |
| C12 | 2.0136 | C42 | 1.5259 | F72 | 0.4274 |
| C13 | 1.7609 | C43 | 1.8506 | F73 | 0.4273 |
| C14 | 1.5794 | C44 | 1.5135 | F74 | 0.3799 |
| C15 | 1.9247 | H45 | 0.3490 | F75 | 0.3793 |
| H16 | 0.3043 | C46 | 1.5855 | F76 | 0.4303 |
| C17 | 1.6888 | C47 | 1.5158 | F77 | 0.4298 |
| H18 | 0.3375 | H48 | 0.3489 | | |
| C19 | 1.7950 | C49 | 1.5818 | | |
| C20 | 1.1814 | C50 | 1.4394 | | |
| H21 | 0.3746 | H51 | 0.4434 | | |
| C22 | 1.1957 | H52 | 0.4084 | | |
| O23 | 1.2841 | C53 | 1.4431 | | |
| O24 | 1.1760 | H54 | 0.4434 | | |
| O25 | 1.1741 | H55 | 0.4084 | | |
| O26 | 1.2831 | H56 | 0.4812 | | |
| N27 | 2.3266 | H57 | 0.4812 | | |
| N28 | 2.3128 | C58 | 1.5107 | | |
| C29 | 1.5292 | H59 | 0.4410 | | |
| C30 | 1.5111 | C60 | 1.5610 | | |

**Figure S17. Atomic polarizability calculation of FDI-2*m*TFPCN**

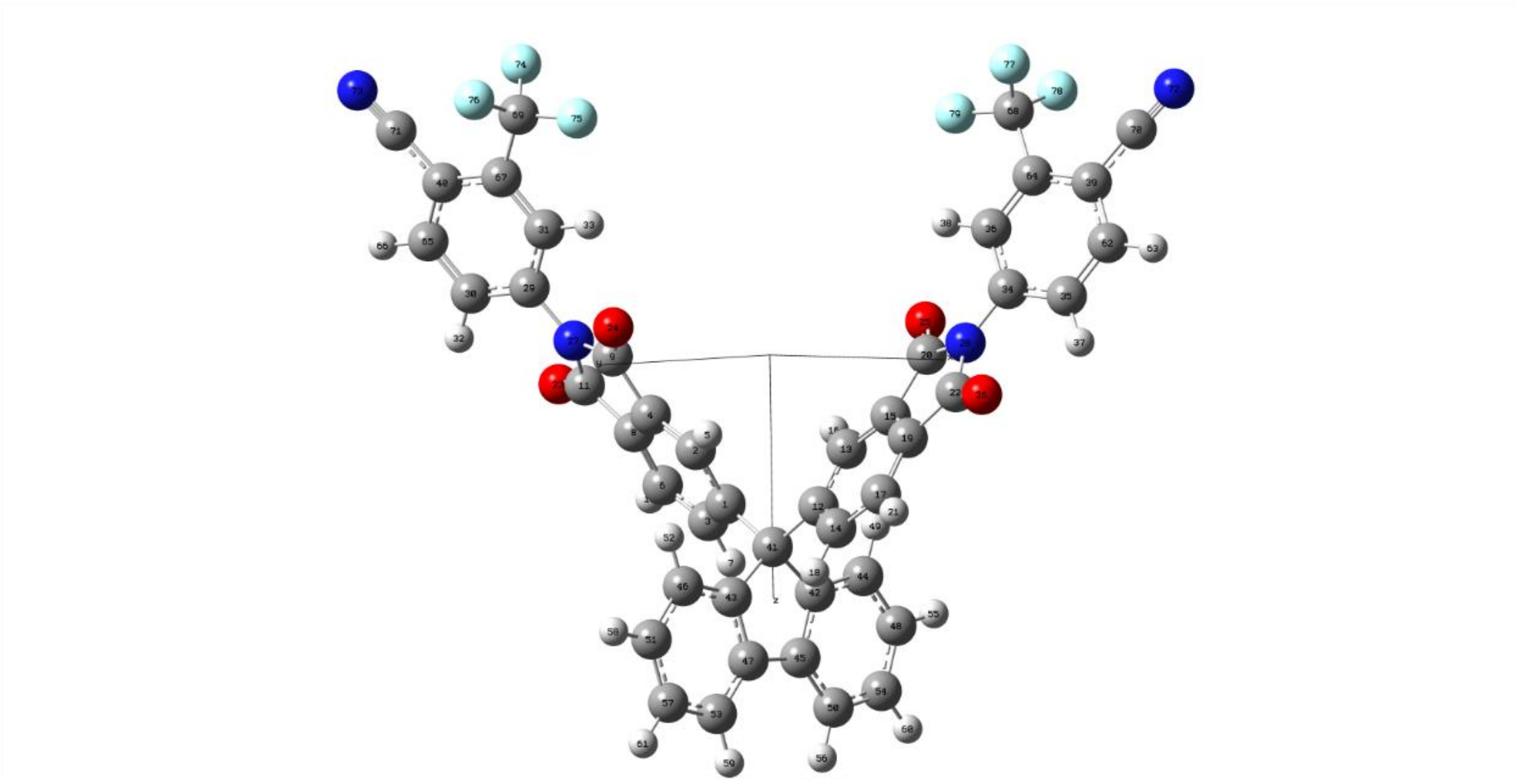

**Figure S23.** Molecular model of FDI-2*m*TFPCN for calculation of atomic polarizabilities.

**Table S8. Atomic polarizabilities of FDI-2*m*TFPCN**

**Table S9.** Calucurated atomic polarizabilities of FDI-2*m*TFPCN.

| | $a_{iso}$ ($10^{-40}$ C m$^2$ V$^{-1}$) | | $a_{iso}$ ($10^{-40}$ C m$^2$ V$^{-1}$) | | $a_{iso}$ ($10^{-40}$ C m$^2$ V$^{-1}$) |
|---|---|---|---|---|---|
| C1 | 2.0812 | C31 | 1.6096 | H61 | 0.4819 |
| C2 | 1.8241 | H32 | 0.4347 | C62 | 1.6183 |
| C3 | 1.6121 | H33 | 0.3942 | H63 | 0.3854 |
| C4 | 2.0054 | C34 | 1.6240 | C64 | 1.7443 |
| H5 | 0.3061 | C35 | 1.6375 | C65 | 1.6245 |
| C6 | 1.7366 | C36 | 1.6072 | H66 | 0.3853 |
| H7 | 0.3407 | H37 | 0.4356 | C67 | 1.7526 |
| C8 | 1.8670 | H38 | 0.3927 | C68 | 0.6269 |
| C9 | 1.2348 | C39 | 1.8184 | C69 | 0.6369 |
| H10 | 0.3729 | C40 | 1.8229 | C70 | 1.4676 |
| C11 | 1.2476 | C41 | 2.7622 | C71 | 1.4696 |
| C12 | 2.0632 | C42 | 1.8274 | N72 | 1.7229 |
| C13 | 1.8122 | C43 | 1.8257 | N73 | 1.7228 |
| C14 | 1.6097 | C44 | 1.5350 | F74 | 0.6307 |
| C15 | 1.9931 | C45 | 1.8688 | F75 | 0.5020 |
| H16 | 0.3059 | C46 | 1.5332 | F76 | 0.6288 |
| C17 | 1.7307 | C47 | 1.8692 | F77 | 0.6309 |
| H18 | 0.3408 | C48 | 1.5256 | F78 | 0.6286 |
| C19 | 1.8608 | H49 | 0.3520 | F79 | 0.3893 |
| C20 | 1.2264 | C50 | 1.5899 | | |
| H21 | 0.3731 | C51 | 1.5237 | | |
| C22 | 1.2434 | H52 | 0.3521 | | |
| O23 | 1.3070 | C53 | 1.5955 | | |
| O24 | 1.2006 | C54 | 1.4472 | | |
| O25 | 1.1993 | H55 | 0.4413 | | |
| O26 | 1.3074 | H56 | 0.4101 | | |
| N27 | 2.4954 | C57 | 1.4455 | | |
| N28 | 2.4937 | H58 | 0.4414 | | |
| C29 | 1.6280 | H59 | 0.4101 | | |
| C30 | 1.6390 | H60 | 0.4818 | | |

**Figure S18. Schematic of dipole orientation in films**

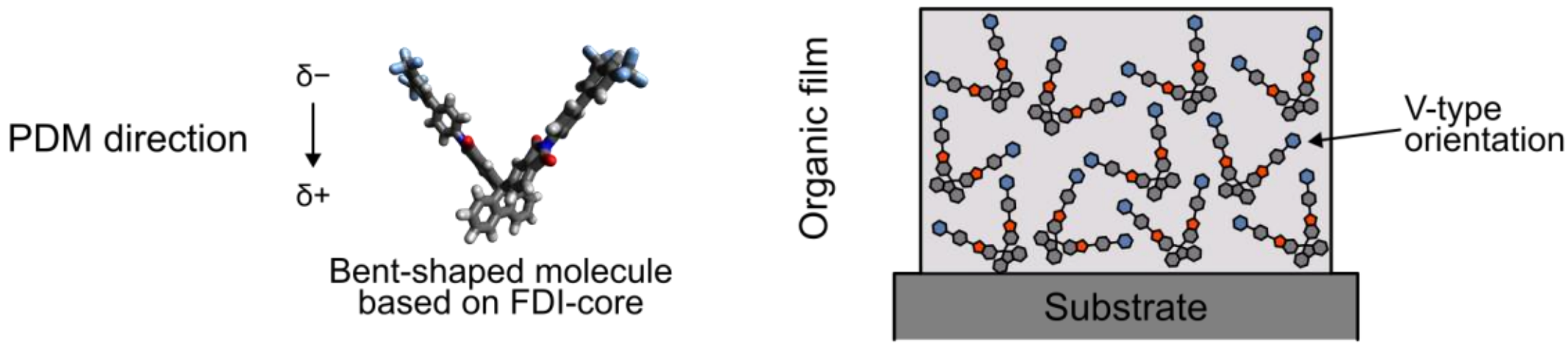

**Figure S18.** Schematics of the spontaneous orientation of polar molecules in films.

**Figure S19. Film capacitance**

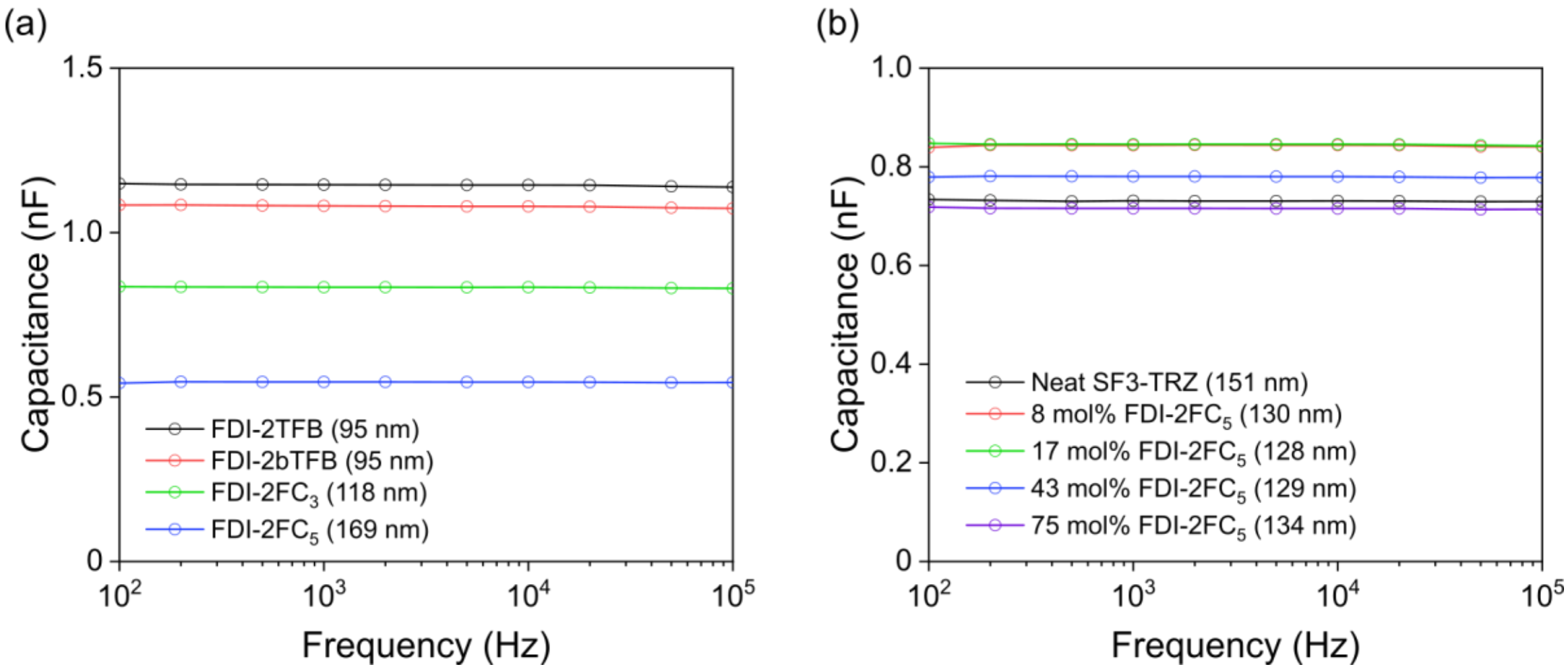

**Figure S19.** C-f curves. (a) The neat film of FDI-based polar molecules. (b) The neat film of SF3-TRZ and doped films of FDI-2FC$_5$:SF3-TRZ.

**Table S9. Relative permittivity of vacuum-deposited films**

**Table S9.** The relative permittivity of the neat and doped films.

| Neat film | $\varepsilon_r$ |
|---|---|
| FDI-2TFB | 3.1 |
| FDI-2bTFB | 2.9 |
| FDI-2FC$_3$ | 2.8 |
| FDI-2FC$_5$ | 2.6 |
| SF3-TRZ | 3.1 |

| Codeposited film with SF3-TRZ | $\varepsilon_r$ |
|---|---|
| 8 mol% FDI-2FC$_5$ | 3.1 |
| 17 mol% FDI-2FC$_5$ | 3.0 |
| 43 mol% FDI-2FC$_5$ | 2.8 |
| 75 mol% FDI-2FC$_5$ | 2.7 |

**Table S10. Film properties**

**Table S10.** Film density and number density.

| | Film density (g cm$^{-3}$) | Number density ($10^{21}$ cm$^{-3}$) |
|---|---|---|
| SO-2PItBu[1] | 1.24 | 1.20 |
| FDI-2TFB | 1.44 | 0.97 |
| FDI-2bTFB | 1.52 | 0.89 |
| FDI-2FC$_3$ | 1.99 | 1.46 |
| FDI-2FC$_5$ | 1.94 | 1.14 |
| SF3-TRZ | 1.40 | 1.54 |

[1] M. Tanaka et al., *Commun. Mater.* 6, 92 (2025).

**Figure S20. SOP properties of FDI-2*p*TFP, FDI-2*m*TFP, and FDI-2*m*TFPCN**

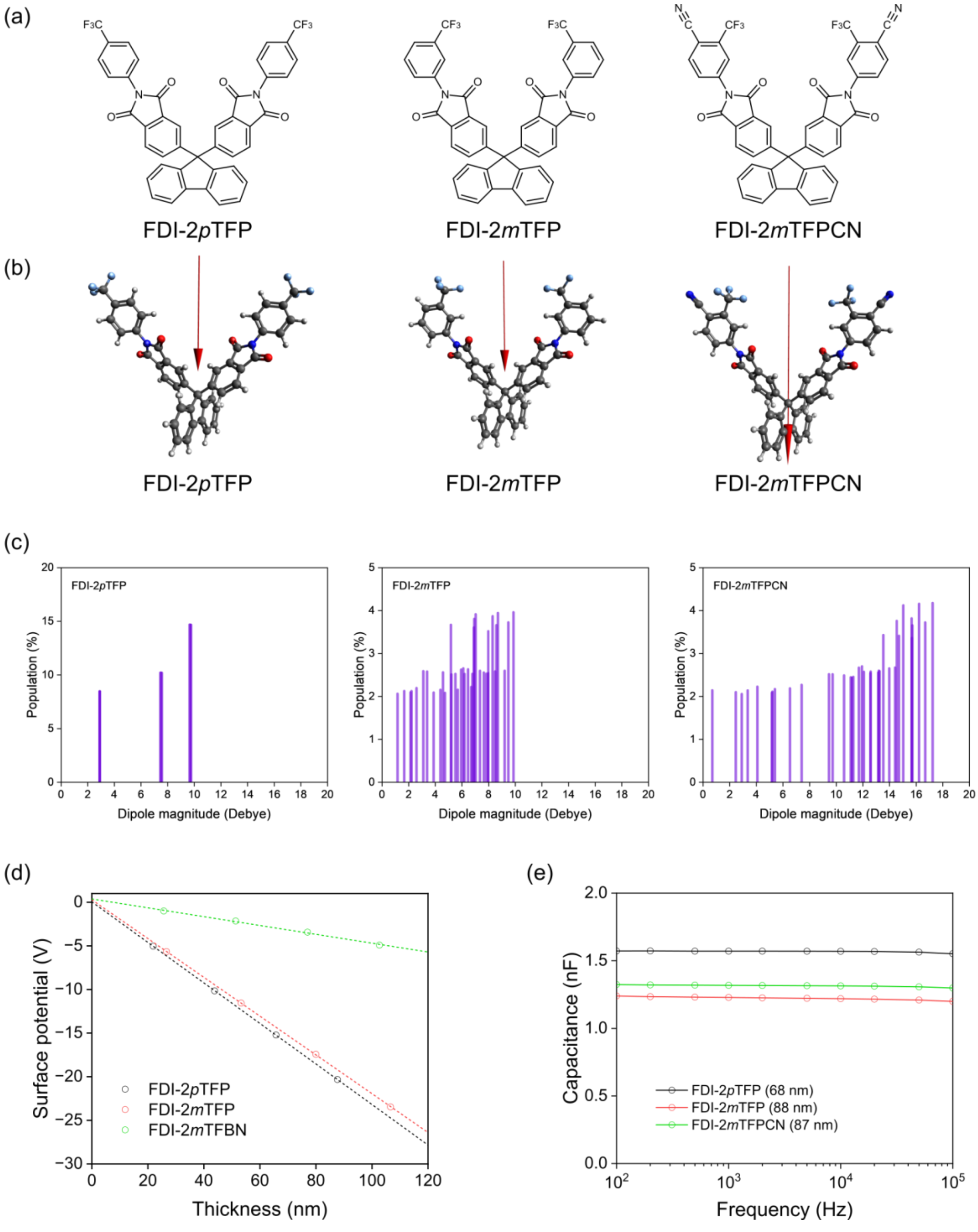

**Figure S20.** SOP properties of FDI-2*p*TFP, FDI-2*m*TFP, and FDI-2*m*TFPCN. (a) Chemical structures of polar molecules. (b) PDM dilections. (c) Distribution of PDM magnitude. (d) Thickness dependence of the surface potential of cvacuum-deposited films. (e) *C-f* curves.

**Table S11. SOP properties of FDI-2*p*TFP, FDI-2*m*TFP, and FDI-2*m*TFPCN**

**Table S11.** Summary of SOP properties of FDI-2*p*TFP, FDI-2*m*TFP, and FDI-2*m*TFPCN.

| | <*p*>[a] (Debye) | GSP slope (mV $nm^{-1}$) | Mean orientation degree | Film density (g $cm^{-3}$) | Number density ($10^{21}$ $cm^{-3}$) | $\varepsilon_r$ |
|---|---|---|---|---|---|---|
| FDI-2*p*TFP | 7.3 | −232 | −0.17 | 1.84 | 1.49 | 3.0 |
| FDI-2*m*TFP | 6.3 | −222 | −0.19 | 1.80 | 1.45 | 3.0 |
| FDI-2*m*TFPCN | 11.7 | −50.7 | −0.028 | 1.72 | 1.30 | 3.2 |

**Figure S21. Stability of surface potential**

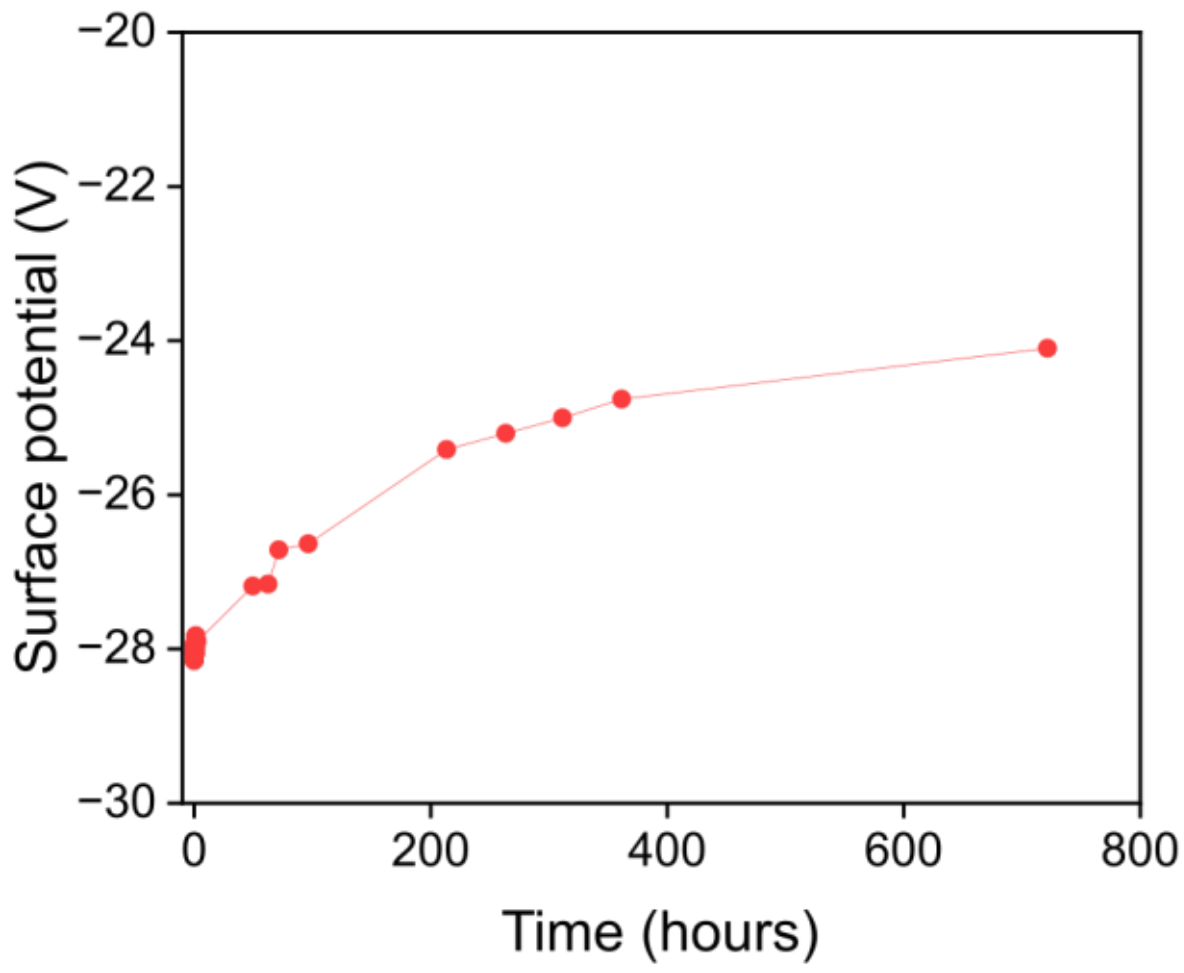

**Figure S21.** Stability of the surface potential of a film of FDI-2FC$_5$. A vacuum-deposited film was stored in the atmosphere under dark conditions.

**Figure S22. Deposition rate dependence of SOP in FDI-2bTFB films**

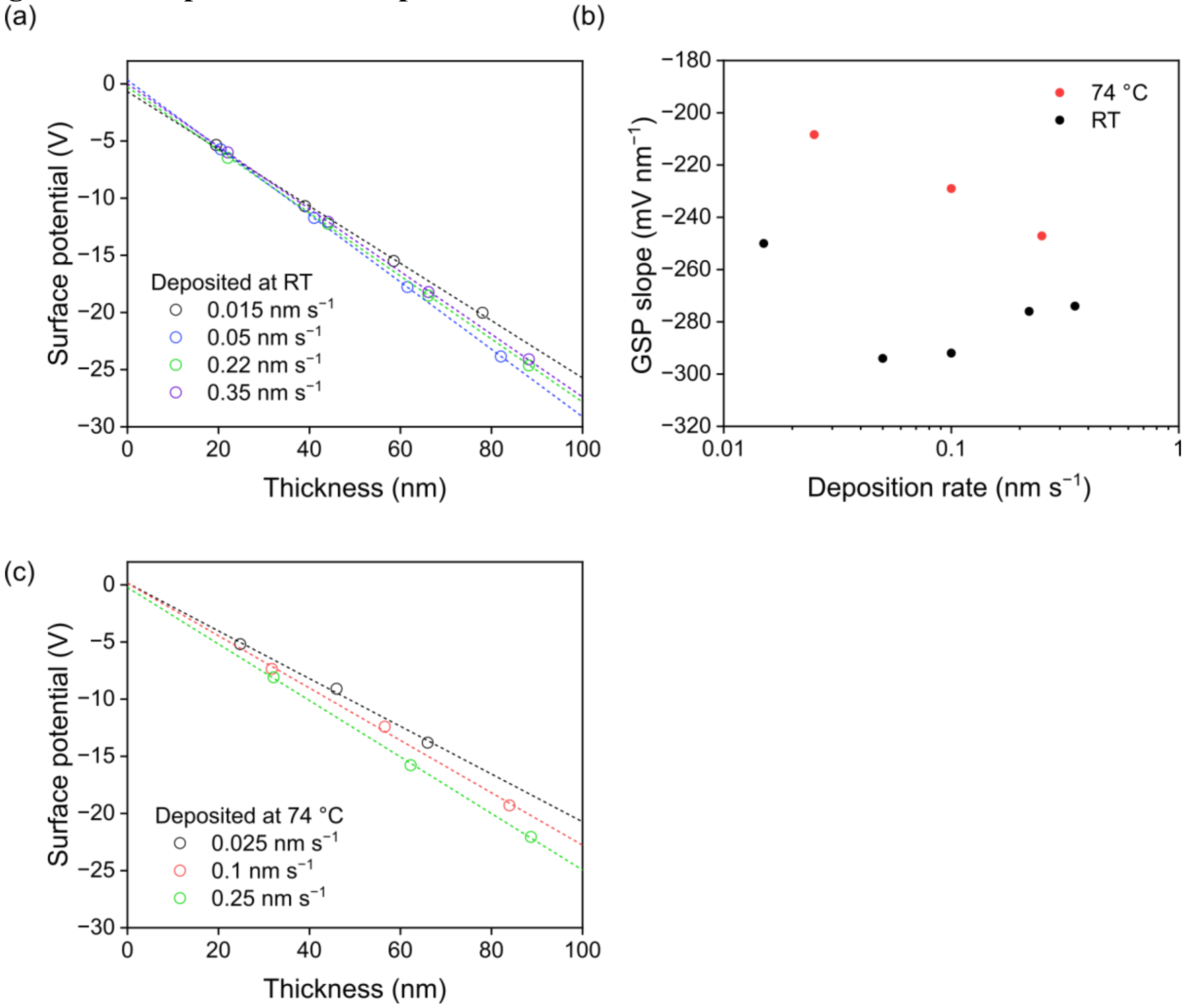

**Figure S22.** Deposition rate dependence of the SOP in FDI-2bTFB films. (a) Thickness dependence of surface potentials of FDI-2bTFB films deposited at $T_s$ of RT. (b) Deposition rate dependence of GSP slopes of FDI-2bTFB films. (c) Thickness dependence of surface potentials of FDI-2bTFB films deposited at $T_s$ of 74 °C.

**Figure S23. DSC results**

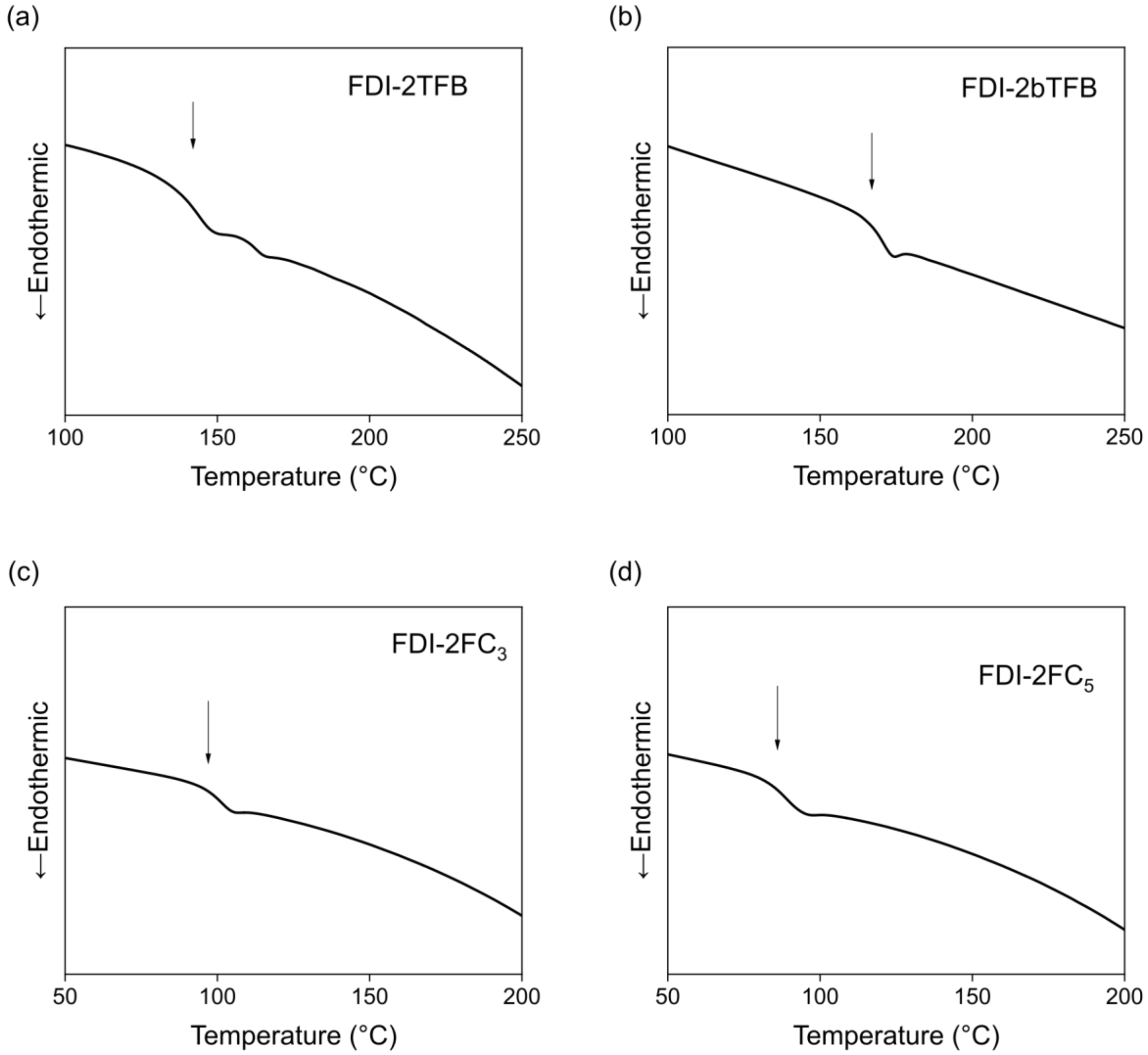

**Figure S23.** DSC results. (a) FDI-2TFB. (b) FDI-2bTFB. (c) FDI-$2FC_3$. (d) FDI-$2FC_5$.

**Figure S24. Nonpolar host molecule**

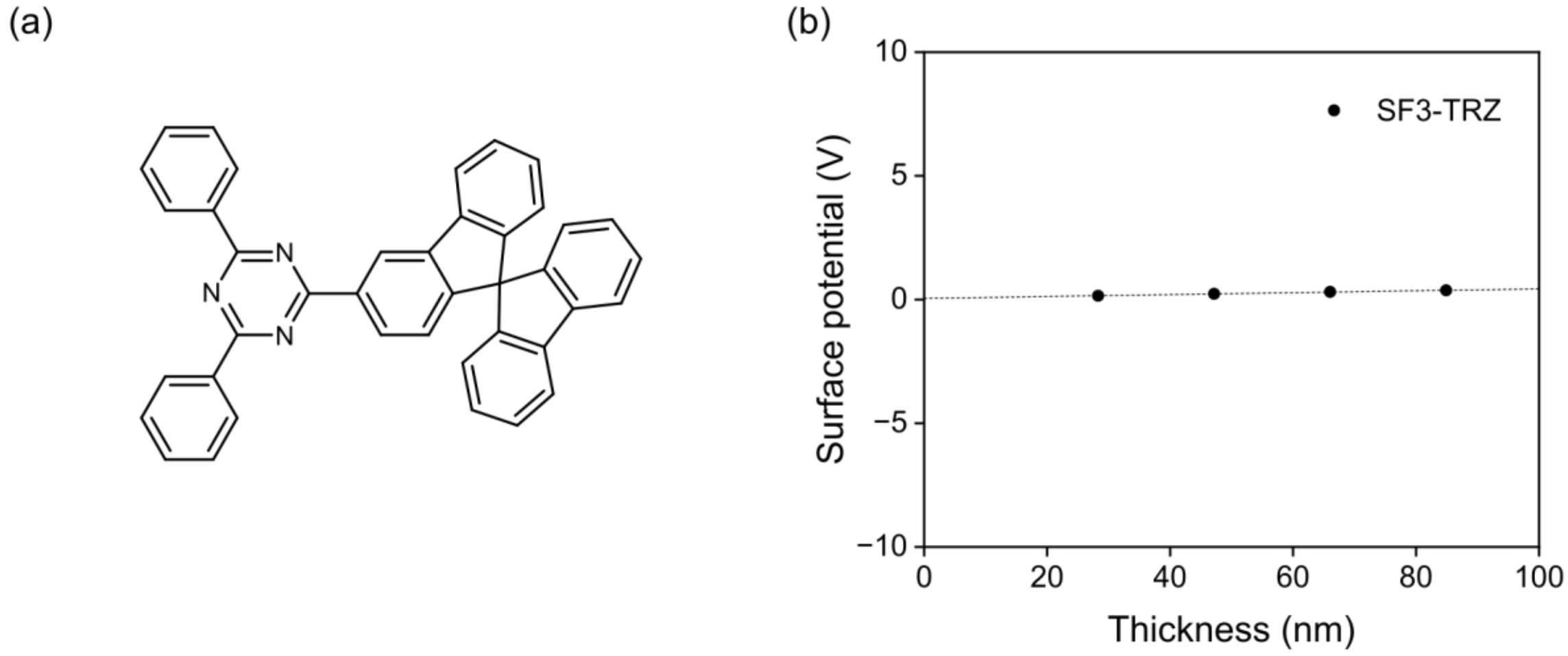

**Figure S24.** Nonpolar host molecule. (a) Chemical structure of SF3-TRZ. (b) Thickness dependence of surface potentials of an SF3-TRZ film.

## Figure S25. Hole-only device

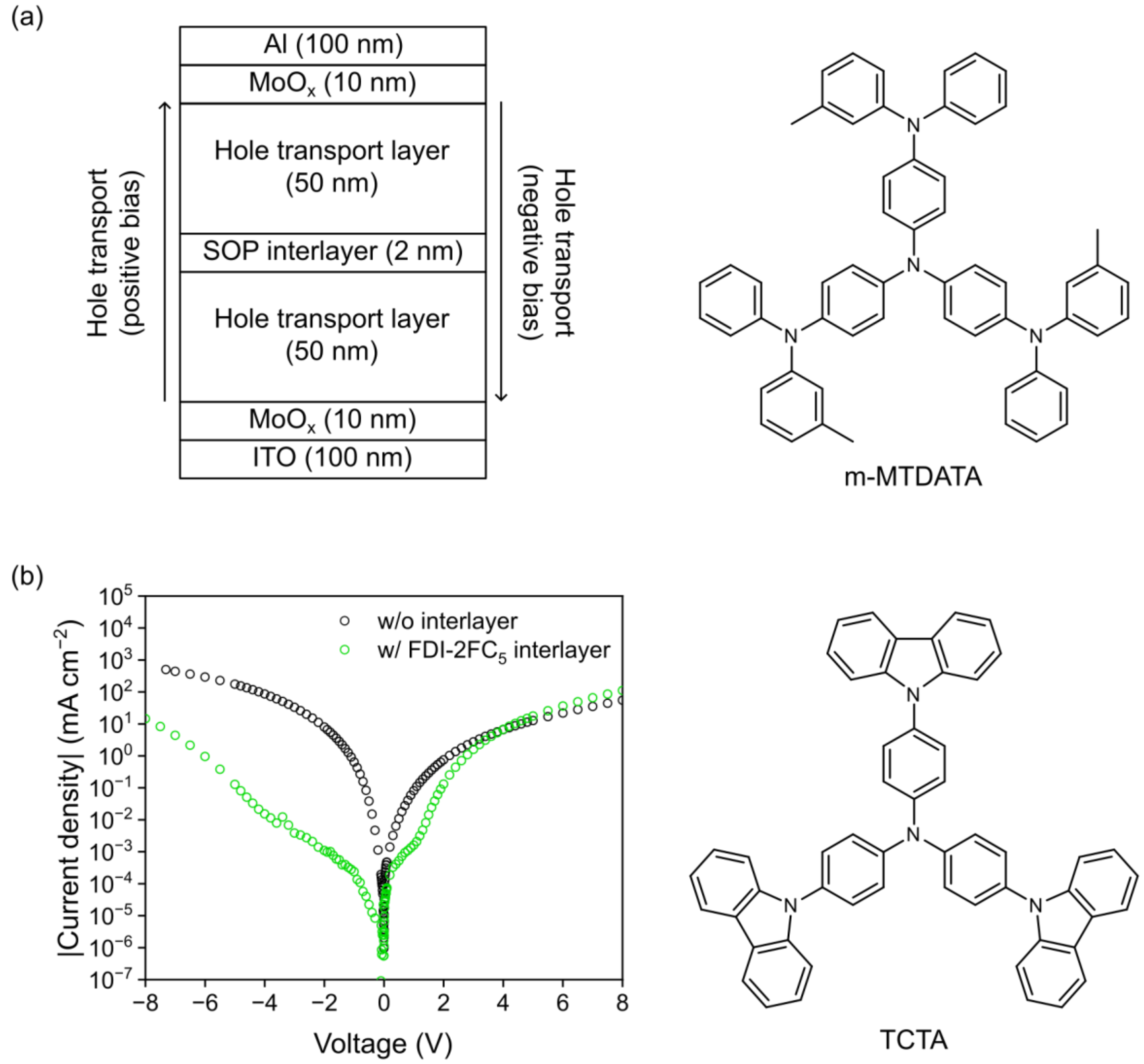

**Figure S25.** Hole-only device. (a) Dvice structure of hole-only device (HOD) amd chemical structure of m-MTDATA. (b) Current density-voltage characteristics of TCTA-based HODs and chemical structure of TCTA.

**Figure S26. HOD based on SO-2PItBu as SOP interlayer**

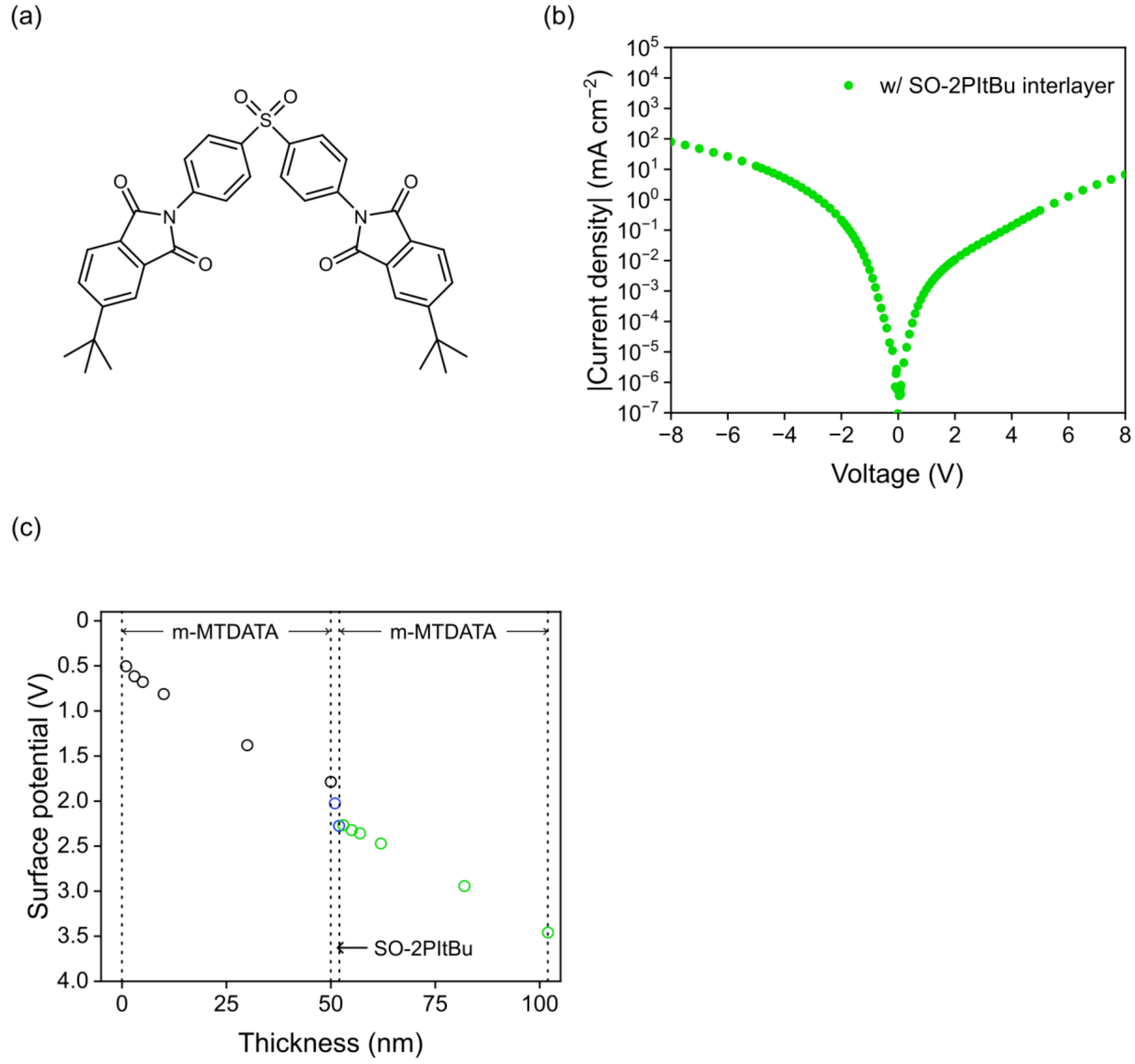

**Figure S26.** HOD based on SO-2PItBu as SOP interlayer. (a) Chemical structure of SO-2PItBu. (b) Current density-voltage characteristics of the m-MTDATA-based HOD with SO-2PItBu interlayer. (c) The thickness dependence of surface potentials of m-MTDATA/SO-2PItBu/m-MTDATA stacks in an ITO substrate.

**Figure S27. PYS and film absorption characteristics**

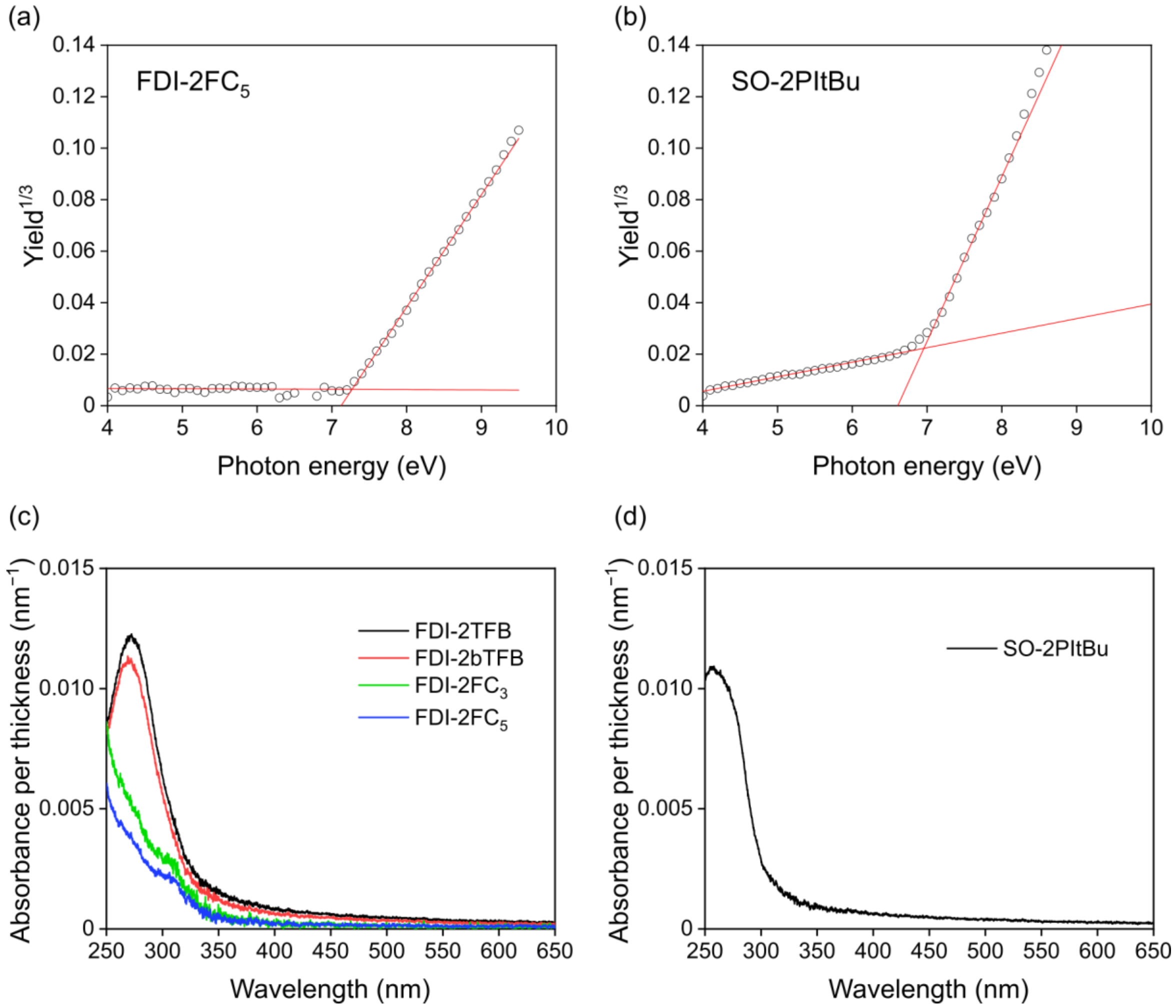

**Figure S27.** Photoemission yield spectroscopy (PYS) and film absorption. (a) PYS result of FSI-2FC$_5$. (b) PYS result of SO-2PitBu. (c) Absorption spectra of FDI-based vacuum-deposited films. (d) Absorption spectrum of an SO-2PitBu film.

**Table S12. HOMO and LUMO levels of SOP molecules**

**Table S12.** HONO and LUMO levels of FDI-2FC$_5$ and SO-2PItBu.

| | HOMO level (eV) | LUMO level (eV) |
|---|---|---|
| FDI-2FC$_5$ | −7.3 | −3.7 |
| SO-2PItBu | −7.0 | −3.2 |

**Figure S28. Energy diagram**

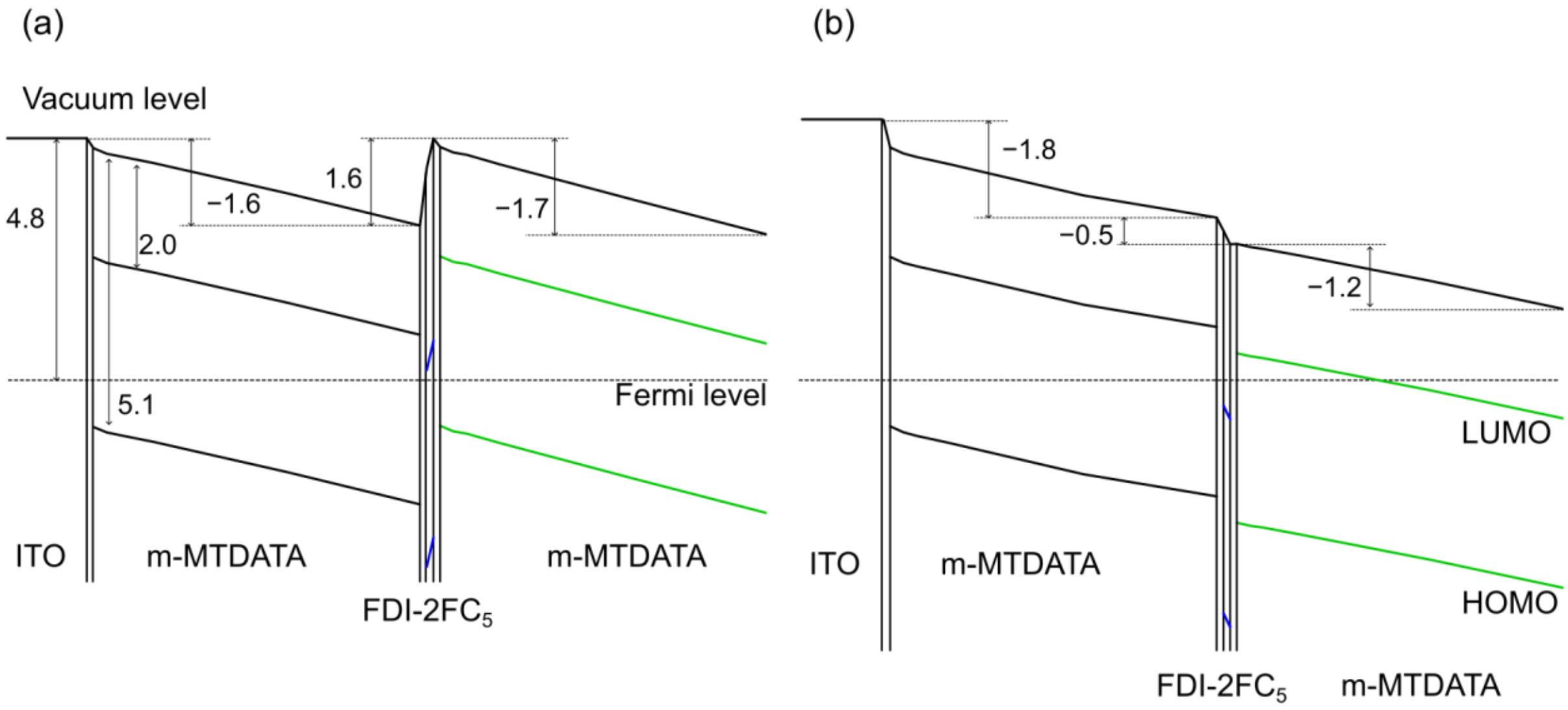

**Figure S28.** Energy diagrams. (a) Energy diagrams of m-MTDATA/FDI-2FC$_5$/m-MTDATA stacks on an ITO substrate. (b) Energy diagrams of m-MTDATA/SO-2PItBu/m-MTDATA stacks on an ITO substrate. These diagrams were derived solely from the results of the surface potential measurements; consequently, the alignment of the Fermi level in the devices is not represented in these diagrams.

**Figure S29. Rectification properties of an HOD with FDI-2FC$_5$ interlayer**

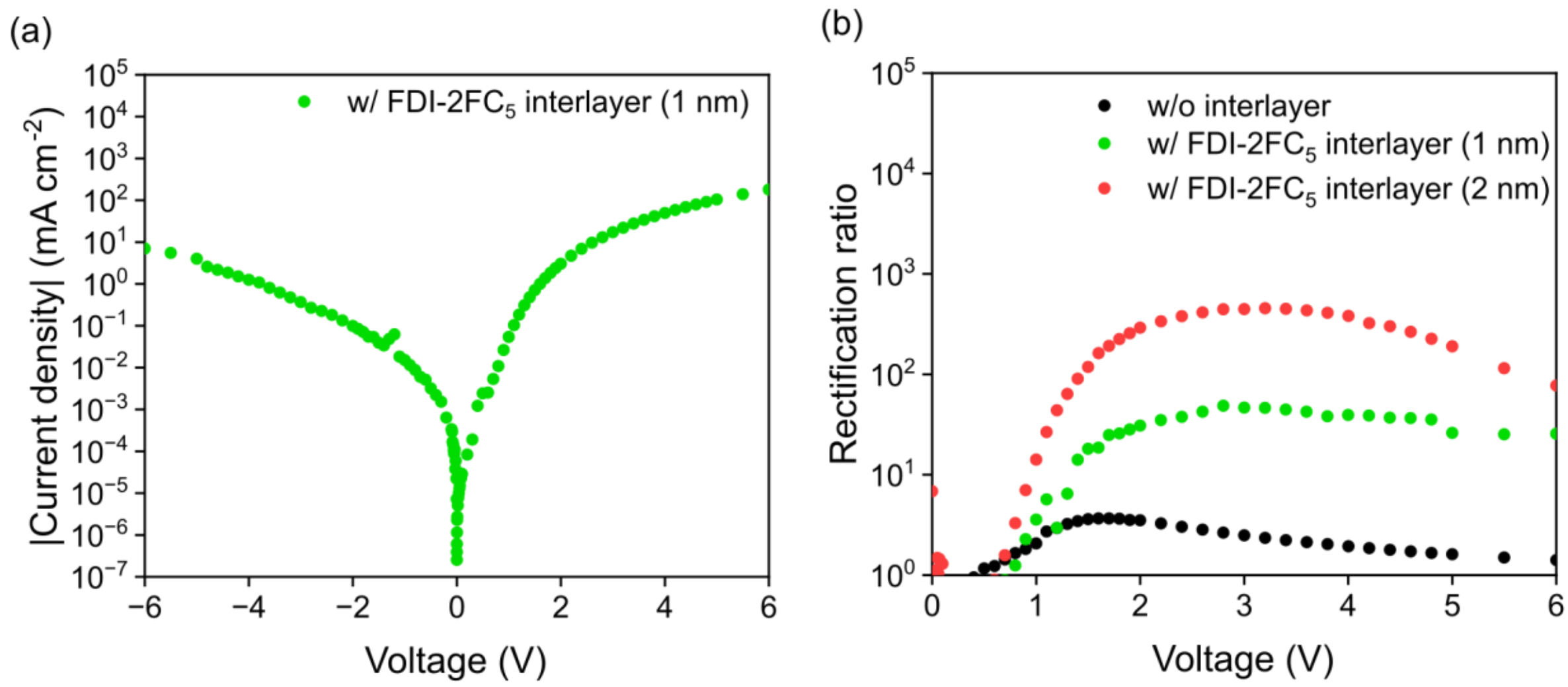

**Figure S29.** Rectification properties of an HOD with FDI-2FC$_5$ interlayer. (a) Current density-voltage characeteristics of an HOD with 1-nm-thick FDI-2FC$_5$ interlayer. (b) Voltage dependence of the rectification ratio of the FDI-2FC$_5$-based HODs with various interlayer thicknesses.

**Figure S30. Surface potential profiles of m-MTDATA/SOP interlayer/$C_{60}$ stacks**

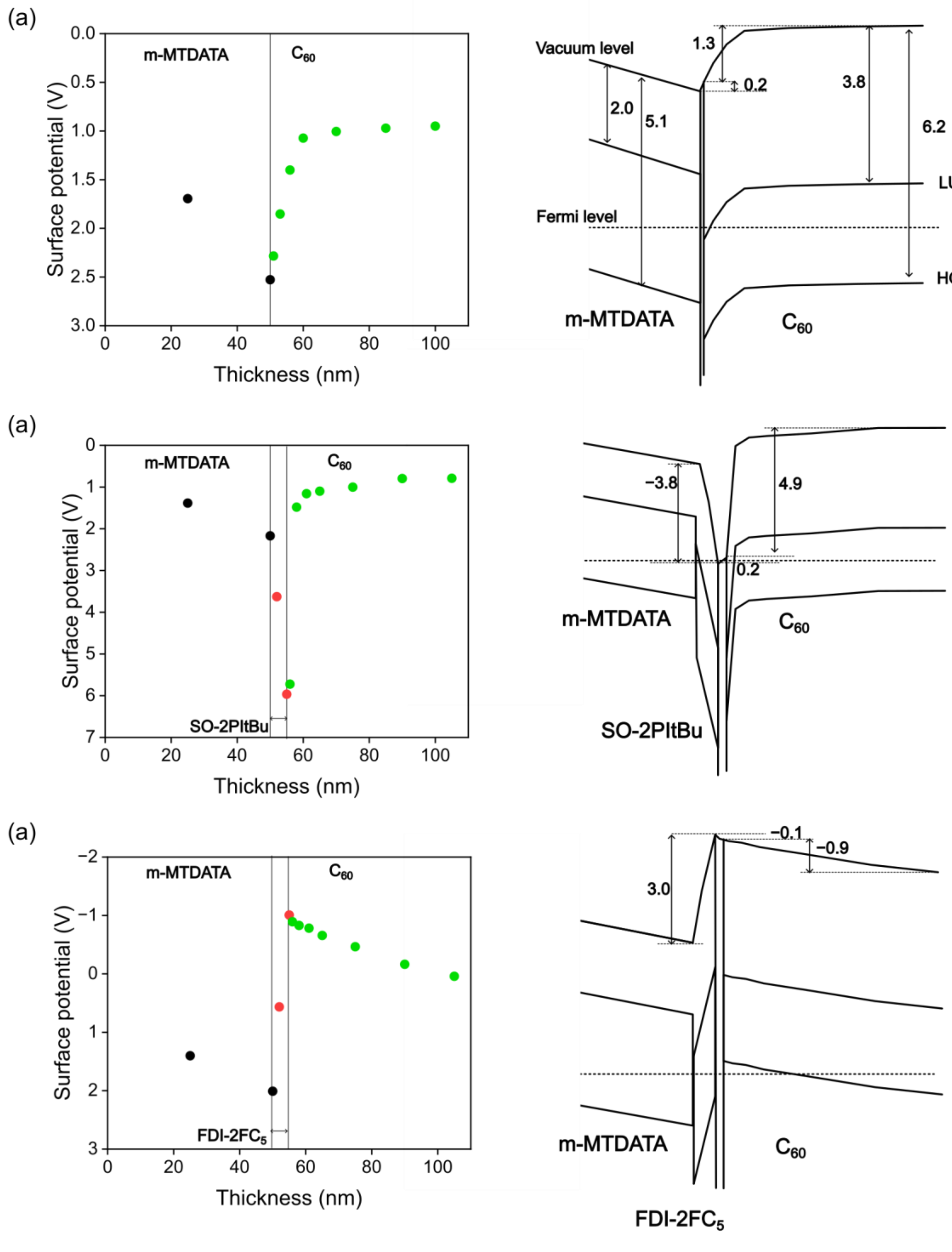

**Figure S30.** Surface potential profiles of m-MTDATA/SOP interlayer/$C_{60}$ stacks on a $MoO_x$/ITO substrate. (a) m-MTDATA (50 nm)/$C_{60}$ (50 nm). (b) m-MTDATA (50 nm)/SO-2PItBu (5 nm)/$C_{60}$ (50 nm). (c) m-MTDATA (50 nm)/FDI-2$FC_5$ (5 nm)/$C_{60}$ (50 nm). Energy diagrams were also depicted. These diagrams were derived solely from the results of the surface potential measurements; consequently, the alignment of the Fermi level in the devices is not represented in these diagrams.

**Figure S31. Photovoltaic properties of OPVs incorporating interlayers under comparable short-circuit current densities**

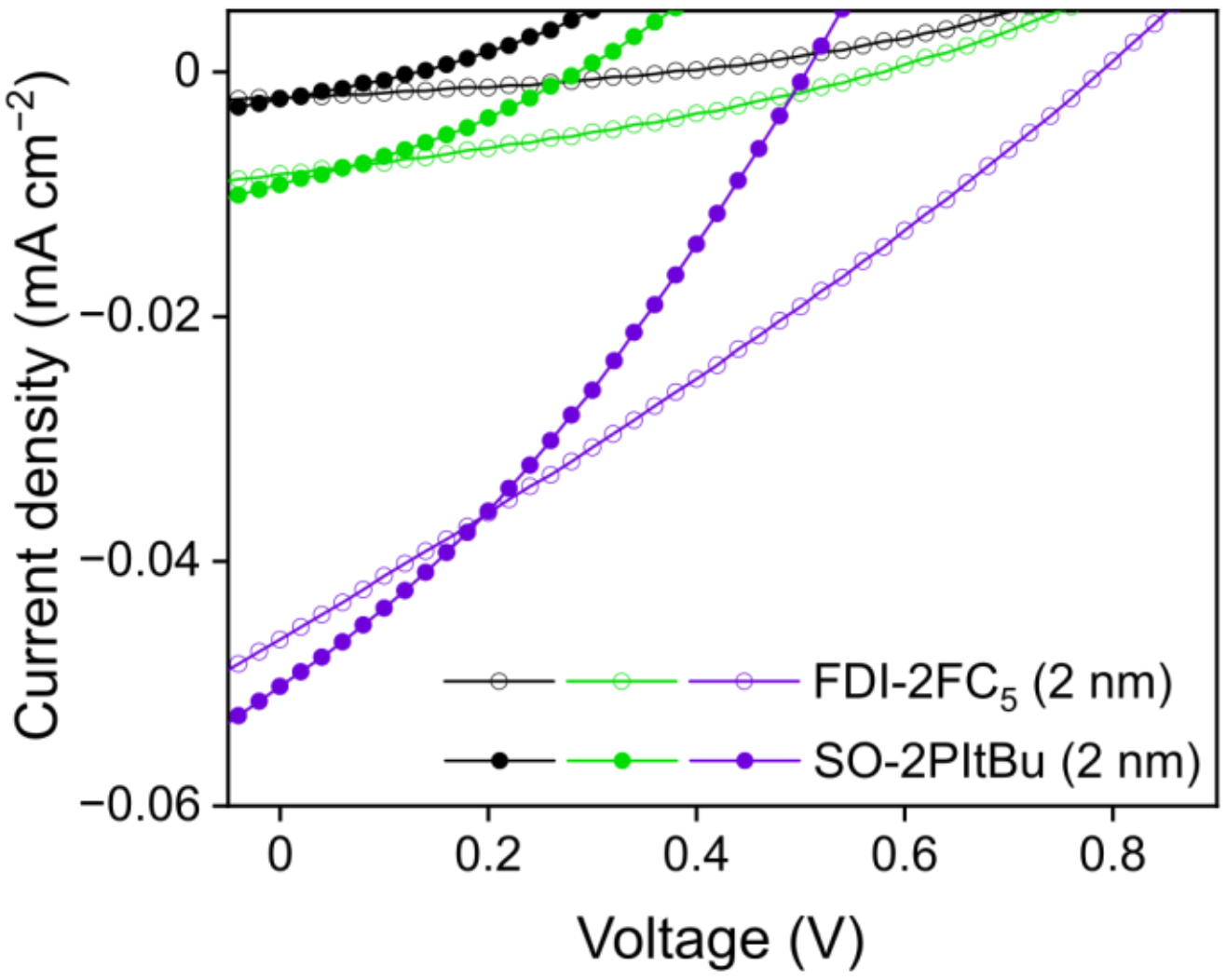

**Figure S31.** Photovoltaic properties of OPVs incorporating interlayers under comparable short-circuit current densities. Open and closed circles show the photovoltaic properties of the OPVs with the interlayers of FDI-2FC$_5$ (2 nm) and SO-2PItBu (2 nm). The intensity of light irradiation to the devices was adjusted to measure various properties with different $J_{SC}$s.

**Figure S32. Photovoltaic properties of B4PyMPM-based OPVs**

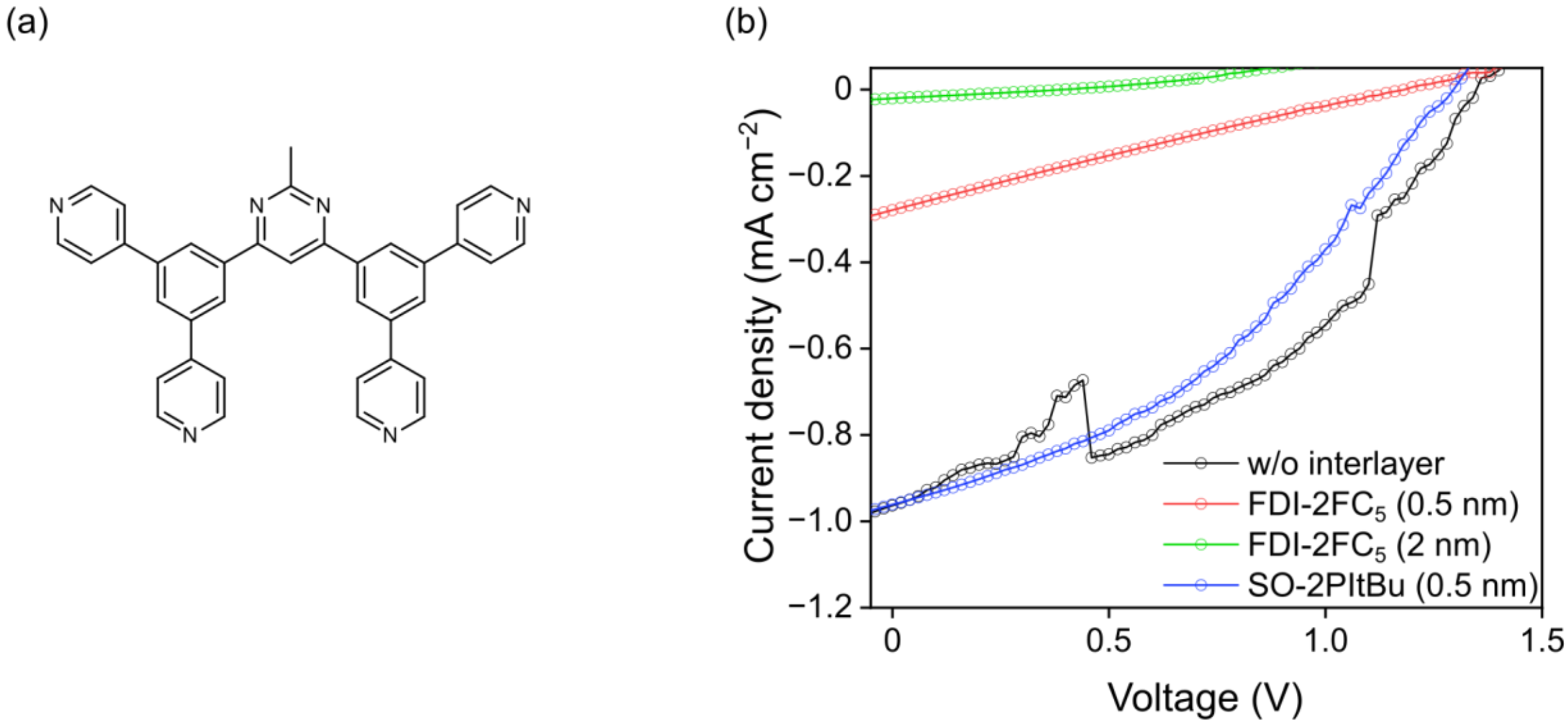

**Figure S32.** Photovoltaic properties of B4PyMPM-based OPVs. (a) chemical structure of B4PyMPM. (b) Current density-voltage characteristics of OPVs. The device structure was ITO (100 nm)/$MoO_x$ (10 nm)/m-MTDATA (50 nm)/SOP interlayer (*x* nm)/B4PyMPM (50 nm)/LiF (1 nm)/Al (150 nm).